\documentclass[3p]{elsarticle}

\usepackage{amssymb}
\usepackage{multirow}
\usepackage{multicol}
\usepackage{graphicx}
\usepackage{amsmath}
\usepackage{lineno}
\usepackage{setspace}
\usepackage{subcaption}
\usepackage{listings}
\usepackage{xspace}
\usepackage{xcolor}
\usepackage{float}
\usepackage{comment}
\usepackage{soul} 

\newcommand{\tripoli}{{\text{Tripoli-4}}\textsuperscript{\textregistered}\xspace}
\newcommand{\mcnp}{\text{MCNP6}\xspace} 
\newcommand{\mcnpver}{\text{MCNP6.2}\xspace}
\newcommand{\geant}{\text{Geant4}\xspace}

\journal{}

\begin{document}

\begin{frontmatter}

\title{Improvement of \geant Neutron-HP package: Unresolved Resonance Region description with Probability Tables}

\author[1,2]{M. Zmeškal}\corref{cor1}
\ead{marek.zmeskal@cvrez.cz}
\author[3]{L. Thulliez}
\ead{loic.thulliez@cea.fr}
\author[4]{P. Tamagno}
\author[3]{E. Dumonteil}

\address[1]{Research Centre Rez, Hlavni 130, Husinec-Rez, 250 68, Czech Republic}
\address[2]{Dept. of Nuclear Reactors, Faculty of Nuclear Sciences and Physical Engineering, Czech Technical University in Prague, V Holesovickach 2, Prague, 180 00, Czech Republic}
\address[3]{Université Paris-Saclay, CEA, Institut de Recherche sur les Lois Fondamentales de l'Univers, 91191, Gif-sur-Yvette, France}
\address[4]{CEA, DES, IRESNE, DER, SPRC, Cadarache, F-13108 Saint-Paul-Lez-Durance, France}

\cortext[cor1]{Corresponding author}

\begin{abstract}

Whether for shielding applications or for criticality safety studies, solving the neutron transport equation with good accuracy requires to take into account the resonant structure of cross sections in part of the Unresolved Resonance Region (URR). In this energy range even if the resonances can no longer be resolved experimentally, neglecting them can lead to significant numerical biases, namely in flux-based quantities. In Geant4, low energy neutrons are transported using evaluated nuclear data libraries handled by the Neutron High-Precision (Neutron-HP) package. In the version 11.01.p02 of the code, the URR can only be described by average smooth cross sections that do not take into account the statistical resonant structure of the cross sections. To overcome this shortcoming, the treatment of the URR with the use of the probability table method has been implemented in Geant4 and successfully validated with the reference Monte Carlo neutron transport codes \mcnp (version 6.2) and \tripoli (version 12). These developments will be taken into account in the next release of \geant. All the validations of \geant have been performed with probability tables generated from both the NJOY and CALENDF pre-processing tools. Therefore \geant now has this unique feature to study the relative impact of the strategies involved during the production of probability table by the two pre-processing codes. This has been used to show that self-shielding is important also for inelastic cross sections in the example of $^{238}$U. The tool to generate probability tables usable by \geant either from  NJOY or from CALENDF is made available on a dedicated GitLab repository and will be included in Geant4.

\end{abstract}

\begin{keyword}

\geant \sep Neutron-HP \sep Neutron \sep Unresolved Resonance Region \sep Probability tables \sep NJOY \sep CALENDF \sep \tripoli \sep \mcnp

\end{keyword}

\end{frontmatter}


\section{Introduction}
\label{sec:introduction}

\begin{figure}[ht!]
    \centering
    \includegraphics[width=0.8\linewidth]{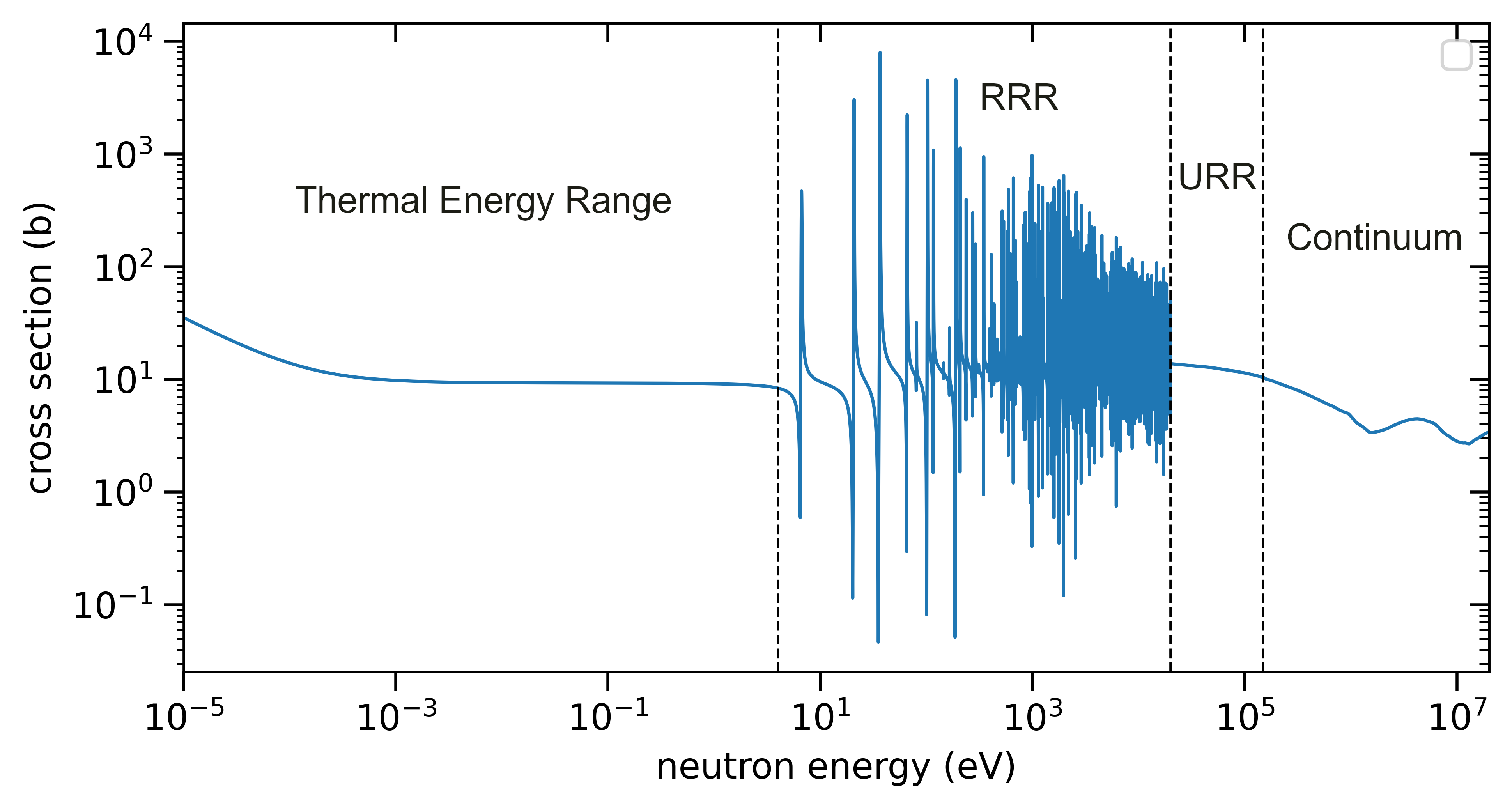}
    \caption{Representation of an usual cross section division in four energy ranges in the example of the elastic cross section of $^{238}$U. In the thermal energy range there is no resonance, in the Resolved Resonance Region (RRR) the resonances are well separated, in the URR (Unresolved Resonance Region) the resonances are not experimentally resolved but still need to be described by statistical distributions and in the continuum region the cross section is described by an optical potential nuclear model.}
    \label{fig:238U_energyRange}
\end{figure}

Monte Carlo codes find widespread use in particle transport applications, enabling the individual tracking of particles and the preservation of correlations between them. To this purpose, \geant has been developed by an international collaboration with the aim of supporting high-energy physics applications as well as other domains such as space science, medical physics, engineering, and nuclear physics~\cite{Agostinelli2003,Allison2006,Allison2016}. In \geant, the individual paths and reactions can be described by nuclear models or by evaluated nuclear data depending on the energy range of the particles. This second approach is used in its Neutron High-Precision (Neutron-HP) package describing low-energy neutrons, \textit{i.e.} below \hbox{20 MeV}, and in reference neutron transport codes such as \tripoli \cite{Brun2015} or \mcnp \cite{Werner2018}. With the increasing needs for neutron transport capabilities within the broader context of multi-particle transport codes, required by applications such as accelerator, medical and fundamental physics in addition to nuclear industry, the Neutron-HP package has recently been under huge improvements \cite{Mendoza2014,Mendoza2018,Thulliez2022, Zmeskal2023} bringing \geant low-energy neutron transport physics almost as precise as reference neutron transport codes such as \tripoli or \mcnp. The last known drawback of the \geant Neutron-HP package is its treatment of cross-sections in the Unresolved Resonance Region (URR), improved in this work by the introduction of probability tables.
\newline
\indent
For low energy neutrons (under 20 MeV), the isotope cross-sections are usually divided into 2 to 4 energy ranges as sketched in Figure \ref{fig:238U_energyRange}, with energy boundaries varying from one isotope to another -- in this case it is described for elastic cross section of $^{238}$U. In the lowest energy part, say below a few eV called the thermal energy range the cross-section does not have any resonances, therefore it is a smooth function of energy. The cross section here is described either with the special thermal scattering law or together with the next region named Resolved Resonance Region (RRR). Then from a few eV to a few keV, the cross-section has huge amplitude variations because of the nuclear states of the compound nucleus formed by the neutron with the target. This is the Resolved Resonance Region (RRR), where the resonances are experimentally well observed, which allows to assign a set of resonance parameters (energy position, width, spin and parity) by fitting the resonance parameters with the help of the R-matrix formalism \cite{Wigner1947}. Above this energy, the resonances can no more be measured (resolved) because the distance between two resonances becomes small compared to the neutron beam experimental energy resolution and because the resonance widths are such that the resonances begin to overlap each other, this is the Unresolved Resonance Region (URR). In the following the URR is further defined as the region where self-shielding is important and needs special treatment in Monte-Carlo particle transport. There, the cross section is obtained from adjusted average resonance parameters. The ENDF files compile the  distribution and average of the partial widths and spacings of the resonances obtained from this adjustment. This region is also a transition between the R-Matrix theory describing the RRR and Hauser-Feshbach theory which describes the smooth cross-section above this region in continuum, in which the individual resonances fully overlap and are no longer distinguishable \cite{Herman2010}. At an energy higher than the URR, the cross section is often described by an optical model potential.

The cross sections are often represented as point-wise tabulated data with prescribed interpolation laws satisfying a given precision reconstruction by a pre-processing code such as PREPRO \cite{Cullen2023} or NJOY \cite{Macfarlane2017}. In the URR, a smooth point-wise average cross section can be used as in \geant -- the only option available until now -- or the so-called probability table (PT) method can be used allowing a better accounting of self-shielding effects \cite{Levitt1972}. Probability tables are usually obtained before the particle transport calculations, also using Monte-Carlo techniques, by aggregation of sampled resonant cross sections. In this latter case, a set of resonances characterizing the cross-section has to be defined. First the resonance energies, \textit{i.e.} their spacing in a given energy range, are sampled from a Wigner distribution knowing the average resonance spacing provided in the ENDF file. Then for each sampled resonance energy, a resonance width is sampled in a $\chi^{2}$ distribution with $N$ degrees of freedom whose parameters are also defined in the ENDF file. This step is repeated for all the spins. Finally the PT is computed by combining all the sampled and reconstructed cross sections contributions. 
Then during a Monte Carlo simulation, when the neutron has an energy inside the URR, the corresponding PT is fetched and a cross-section is randomly sampled from it. 
\newline
Taking into account resonances in the URR by means of a PT is important for example for criticality studies since the use of PTs can increase the reactivity by up to 500 pcm for fast and medium systems with low enrichment due to self-shielding effect from $^{238}$U \cite{Weinman1998, Mosteller1999, Walsh2017} and also for shielding applications since the dose rate behind a $^{238}$U shielding can increase by up to a factor $2.42$ \cite{Carter1998}. Thus omitting the statistical resonance structure in the URR could lead to non-conservative results. 
\newline
Until now, \geant has only used smooth average cross sections to describe the URR. Therefore this paper reports the implementation of the URR within the \geant Neutron-HP package. The benchmark methodology and the associated tools used in this work are detailed in Section \ref{sec:methodology}. Then the different pre-processing codes, along with their specificities, used to build probability tables for \geant are presented in Section \ref{sec:probTabs} along with a description of the technical implementation of the PT in \geant. In Section \ref{sec:validations} the validation of the URR description by the PT method in \geant is performed with the reference neutron transport codes \tripoli and \mcnp. Finally the influence of the choice of the NJOY or CALENDF pre-processing tool, used to generate the PT, on Monte Carlo simulation results is studied.

\section{Benchmark methodology and tools}
\label{sec:methodology}

This study has been performed with \geant version 11.01.p02. The  developments regarding the use of PT have been validated against the reference neutron transport codes \tripoli~(version 12) and \mcnp~(version~6.2). \tripoli is developed since the mid-1990s at CEA-Saclay (France) and is used as a reference code by the main French nuclear companies \cite{Brun2015}. \mcnp is developed at Los Alamos National Laboratory (United States) and is used worldwide as a reference neutron transport code for many applications involving neutrons \cite{Werner2018}. Both codes are used for criticality-safety studies, reactor physics and shielding applications. They benefit from a very large qualification database gathering more than 1000 experimental benchmarks and are regularly validated using inter-code comparisons \cite{intercomp}.
\newline
The evaluated nuclear data library ENDF/B-VII.1 \cite{Chadwick2011} is used in this work.
\newline
In the following, the developments performed in \geant are validated with a macroscopic benchmark that will be referred to as the sphere benchmark. It consists of a homogeneous sphere with a $50$ cm radius from a given material -- $^{238}$U or tungsten isotopes -- in which the neutron flux is recorded using a track length estimator. A 1 cm thick outer shell made of vacuum is used to record the outgoing flux, still with a track length estimator. An isotropic point-like neutron source is placed at its center. Two types of initial neutron energy spectra are used. The first one is a 1/$E$ energy spectrum from $10$ keV to $300$ keV used for the $^{238}$U isotope allowing to simulate a typical reactor spectrum. The second one is a uniformly distributed energy spectrum from $1$ keV to $200$ keV for the tungsten isotopes, mainly used in shielding applications. The energy ranges of the input energy spectra have been chosen to cover the whole URR range of the studied isotopes as shown in Table \ref{tab:material}. This benchmark is an adaptation of the already existing benchmarks that can be found on the following GitLab repository \cite{MarekGitLab2023}. Each simulation has been performed with 10$^{9}$ initial neutrons and normalized to one source neutron per second. The results of two codes are compared relative to three times the statistical uncertainties (3$\sigma_{\text{stat}}$).
\newline
Hereafter the validations of the use of PT generated by either NJOY \cite{Macfarlane2017} or CALENDF \cite{Sublet2011} are respectively performed with \mcnp and \tripoli. It has to be pointed out that for these benchmarks the exact same probability tables are always used. \geant and \mcnp are compared with PT from NJOY and \geant and \tripoli with PT from CALENDF, following the usual calculation scheme of both reference codes. It is worth noticing that \geant is here the only code able to deal with both NJOY and CALENDF probability tables. 
\newline
In \geant the Doppler broadening of the 0 K cross sections can be activated or deactivated in using the command line \hbox{``/process/had/particle\_hp/neglect\_Doppler\_broadening true/false''}. If it is deactivated, \geant uses the 0 K cross sections. When it is activated the Doppler broadening is performed stochastically on-the-fly, at any temperature, allowing to accommodate the different physics fields requirements. However this comes at a huge computational cost. For the physics cases dealt with in this work, this prevents to perform the simulations in a decent amount of time, \textit{i.e.} in less than 30 days. Therefore in the following, a workaround has been used, which is now explained. Pre-Doppler broadened cross sections have been prepared at 294 K with the BROADR module of NJOY with a reconstruction tolerance of 0.1 \%. After that, the 0 K cross sections have been replaced by the newly generated 294 K ones in the \geant cross section database. Then in the simulation the on-the-fly Doppler broadening is deactivated resulting in the use of the pre-Doppler broadened cross sections at 294 K. This workaround leads to decrease of the computational time by at least a factor 100 for $^{238}$U. It has to be mentioned that both \mcnp and \tripoli usually use pre-Doppler broadened cross sections. The Doppler Broadening Rejection Correction (DBRC) option is turned off in \tripoli \cite{Zoia2013} and \geant \cite{Zmeskal2023}, it is not available in \mcnpver.

\begin{table}[h!]
\centering
\caption{Characteristics of the different sphere benchmarks used in this work. The URR energy limits of each isotope are from ENDF/B-VII.1. The temperatures are in K, the densities in g/cm$^3$, the abundances in \% and the energies in keV.}
\begin{tabular}{c|cc|cc|cc|cc}
Material  & Temperature & Density & \multicolumn{2}{c|}{Energy spectrum} & Isotopes  & Abundance & \multicolumn{2}{c}{URR energy limits} 
\\ 
   &     &   & Type & Energy range & & &  Min & Max
\\ \hline
$^{238}$U    & 293.15     & 18.9    & 1/$E$ & 10-300      & $^{238}$U & 100    & 20 & 149 
\\ \hline
$^{184}$W    & 294        & 19.25   & flat & 1-200     & $^{184}$W & 100    & 4 & 100    
\\ \hline
\multirow{5}{*}{$^{\text{nat}}$W} & \multirow{5}{*}{294} & \multirow{5}{*}{19.25}   & \multirow{5}{*}{flat}  & \multirow{5}{*}{1-200 } 
& $^{180}$W & 0.12   & - & -      
\\
&   &   &    &    & $^{182}$W & 26.50  & 4.5 & 100                                             
\\
 &  &   &    &    & $^{183}$W & 14.31  & 2.2 & 45                                              
\\
&   &   &    &    & $^{184}$W & 30.64  & 4.0 & 100                                             
\\
&   &   &    &    & $^{186}$W & 28.43  & 8.5 & 100                                            
\end{tabular}
\label{tab:material}
\end{table}

\section{Probability tables in \geant}
\label{sec:probTabs}

\subsection{NJOY and CALENDF processing tools}

The two processing codes used in this work to generate probability tables are NJOY and CALENDF.   
\mcnp uses probability tables generated by the PURR module of NJOY written in the ACE format, while \tripoli uses the one from CALENDF. In evaluated nuclear data files, reactions are labeled by an MT number, for instance MT=1 for total, MT=2 for elastic scattering, \textit{etc}. The related partial cross sections in CALENDF are however gathered for some usages as macro-reaction cross sections (MTREP). In this work, a special attention has been paid to the partial cross section grouping in CALENDF as presented in Table~\ref{tab:calendf_njoy_mt} to agree with their Geant4 handling.  Although both, NJOY and CALENDF, codes create probability tables with the same approach of the sampling from ladders, there are some differences that could give rise to discrepancies, which are now discussed.

\begin{table}[h!]
    \centering
    \caption{Cross section grouping and treatment performed in CALENDF and NJOY to generate probability tables for \geant.}
    \begin{tabular}{c|c|c|c|c}
                & MT=2    & MT=102   & MT=18   & All the others MTs \\
        \hline
       CALENDF  &  PT (MTREP=1) & PT (MTREP=2)  & PT (MTREP=3) & PT (MTREP=4+5) \\
        \hline
       NJOY     &  PT            & PT             &  PT     & Smooth    \\
    \end{tabular}    
    \label{tab:calendf_njoy_mt}
\end{table}

\begin{figure}[h!]
    \centering
    \includegraphics[width=0.8\linewidth]{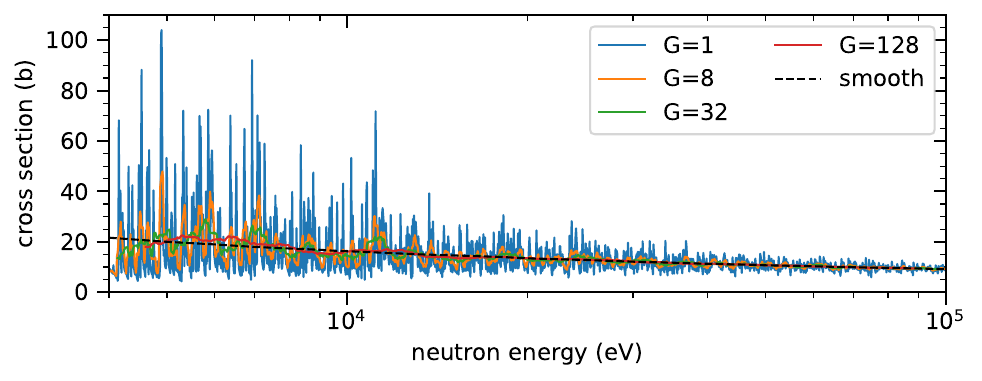}
    \caption{Elastic cross section of $^{184}$W in the URR energy range between 4 keV and 100 keV computed with CALENDF with ipreci=4. A moving average algorithm is used with different energy group width $G$ and the results are compared to the smooth cross section.}
    \label{fig:w184groups}
\end{figure}

\begin{figure}[h!]
    \centering
    \includegraphics[width=0.8\linewidth]{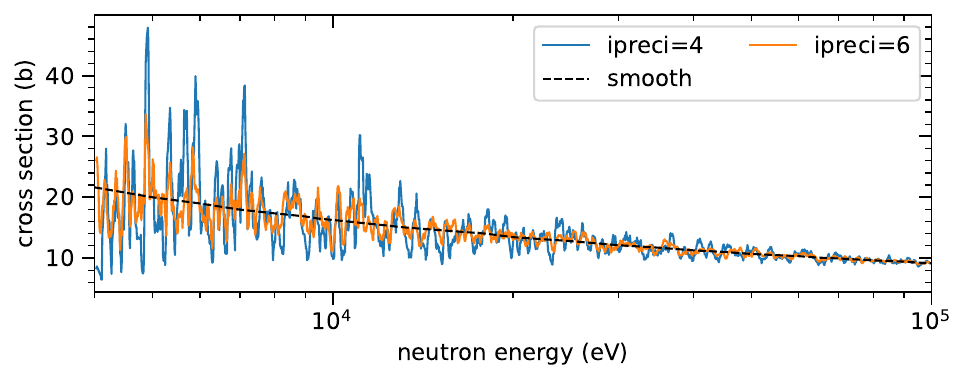}
    \caption{Elastic cross section of $^{184}$W in the URR energy range between 4 keV and 100 keV computed with CALENDF with ipreci=4 and ipreci=6. The results are plotted with a moving average algorithm and an energy group width $G=8$ along with the smooth cross section.}
    \label{fig:w184ipreci}
\end{figure}

\begin{figure}[h!]
     \centering
     \begin{subfigure}[b]{0.49\textwidth}
         \centering
         \includegraphics[width=\textwidth]{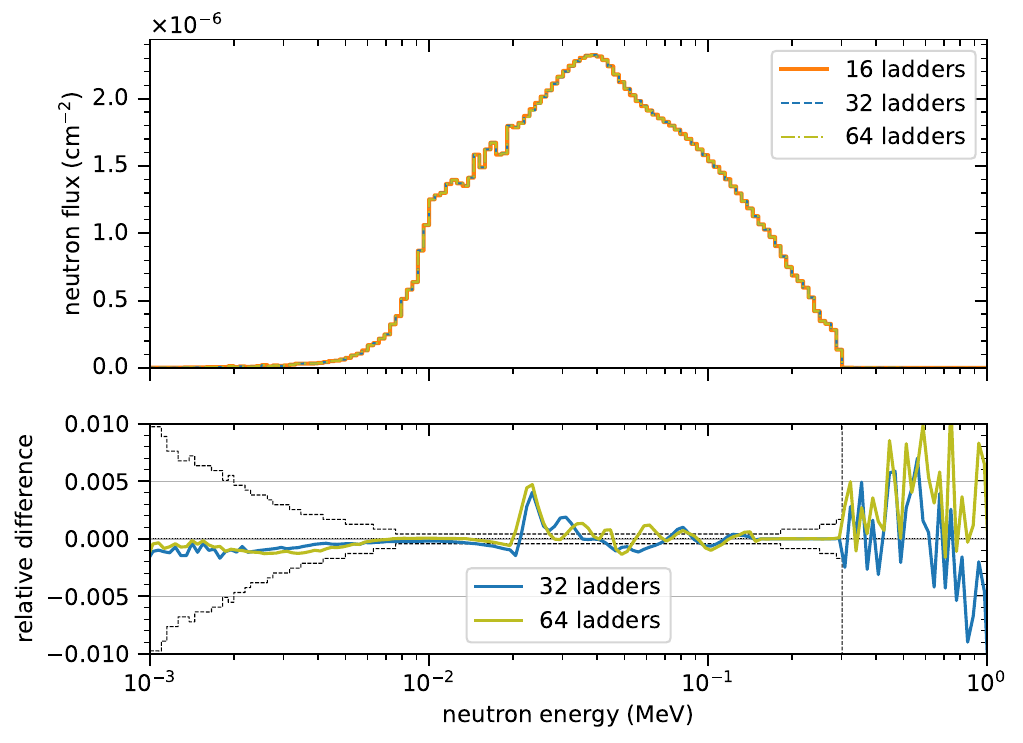}
         \caption{NJOY - $^{238}$U - \mcnp}
         \label{fig:laddersU}
     \end{subfigure}
     \hfill
     \begin{subfigure}[b]{0.49\textwidth}
         \centering
         \includegraphics[width=\textwidth]{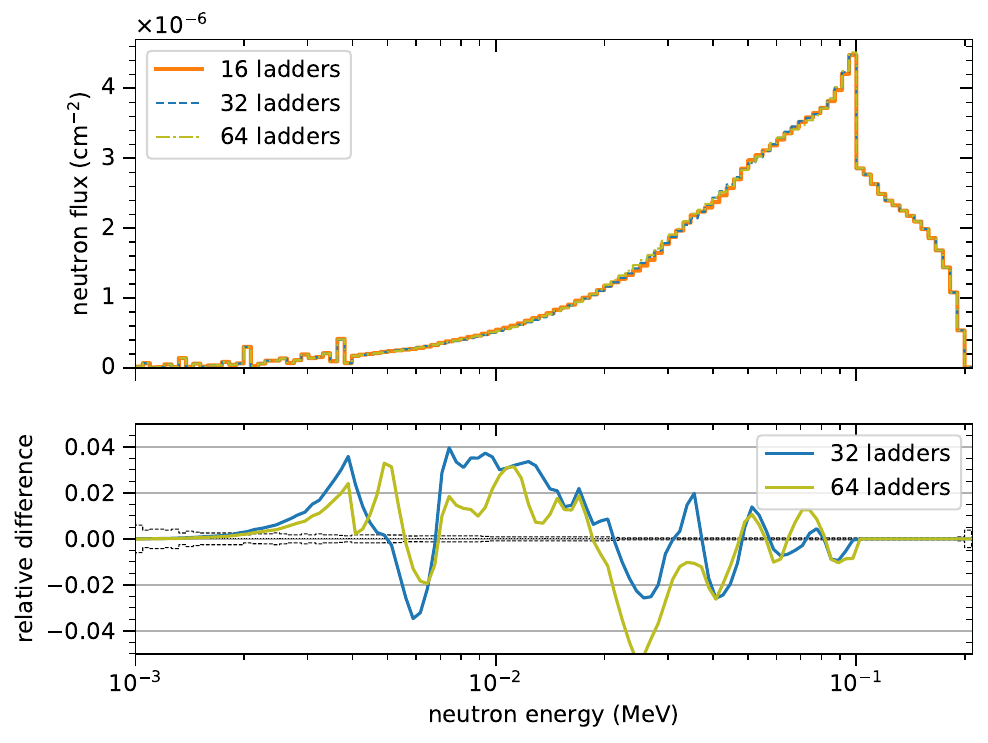}
         \caption{NJOY - $^{184}$W - \mcnp}
         \label{fig:laddersW}
     \end{subfigure}
     \begin{subfigure}[b]{0.49\textwidth}
         \centering
         \includegraphics[width=\textwidth]{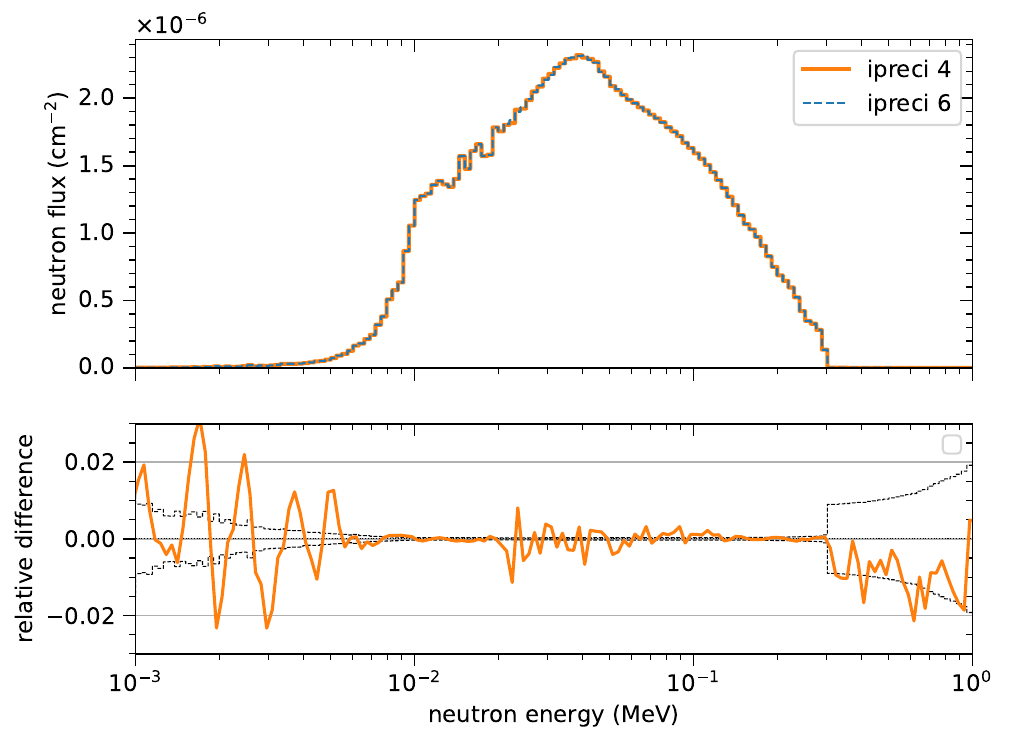}
         \caption{CALENDF - $^{238}$U - \geant}
         \label{fig:laddersW184}
     \end{subfigure}
      \hfill
     \begin{subfigure}[b]{0.49\textwidth}
         \centering
         \includegraphics[width=\textwidth]{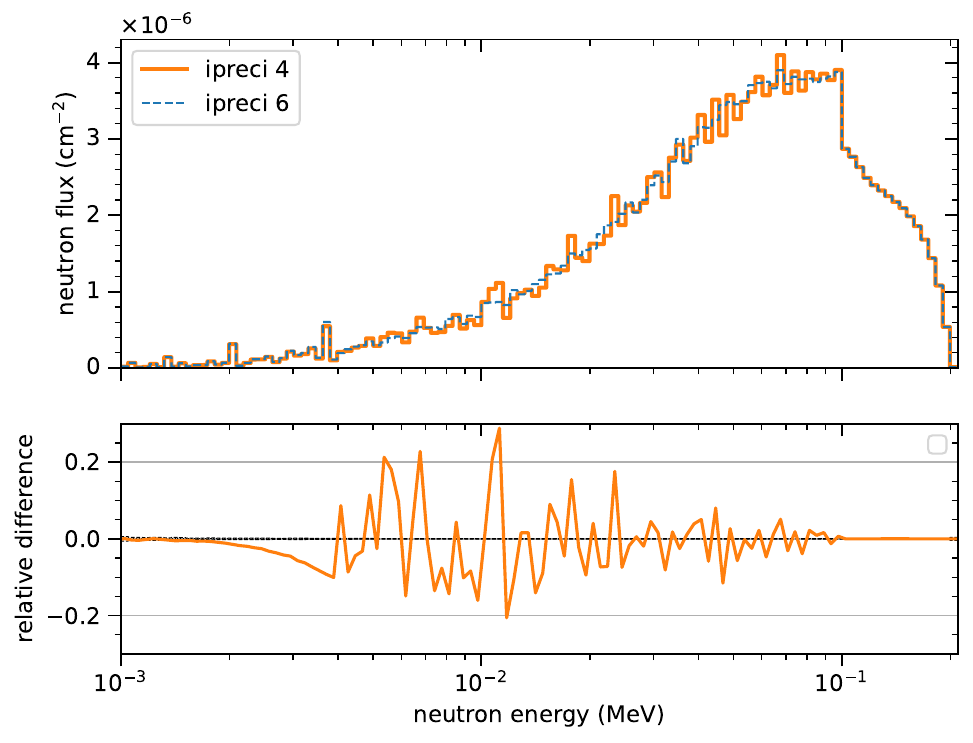}
         \caption{CALENDF - $^{184}$W - \geant}
         \label{fig:ipreciW184benchmark}
     \end{subfigure}
        \caption{\textbf{Top:} Influence of the number of ladders (16, 32 or 64) used in NJOY to generate the probability tables on the $^{238}$U and $^{184}$W benchmarks performed with \mcnp. \textbf{Bottom:} Influence of the number of samples in a given energy bin (ipreci=4 for 4 samples and ipreci=6 for 16 samples) used in CALENDF to generate the probability tables on the $^{238}$U and $^{184}$W benchmarks performed with \geant. In the relative difference plots, the 3$\sigma_{\text{stat}}$ uncertainties are represented by the dashed lines.}
    
        \label{fig:ladders}
\end{figure}

\begin{figure}[!h]
         \centering
         \includegraphics[width=0.8\linewidth]{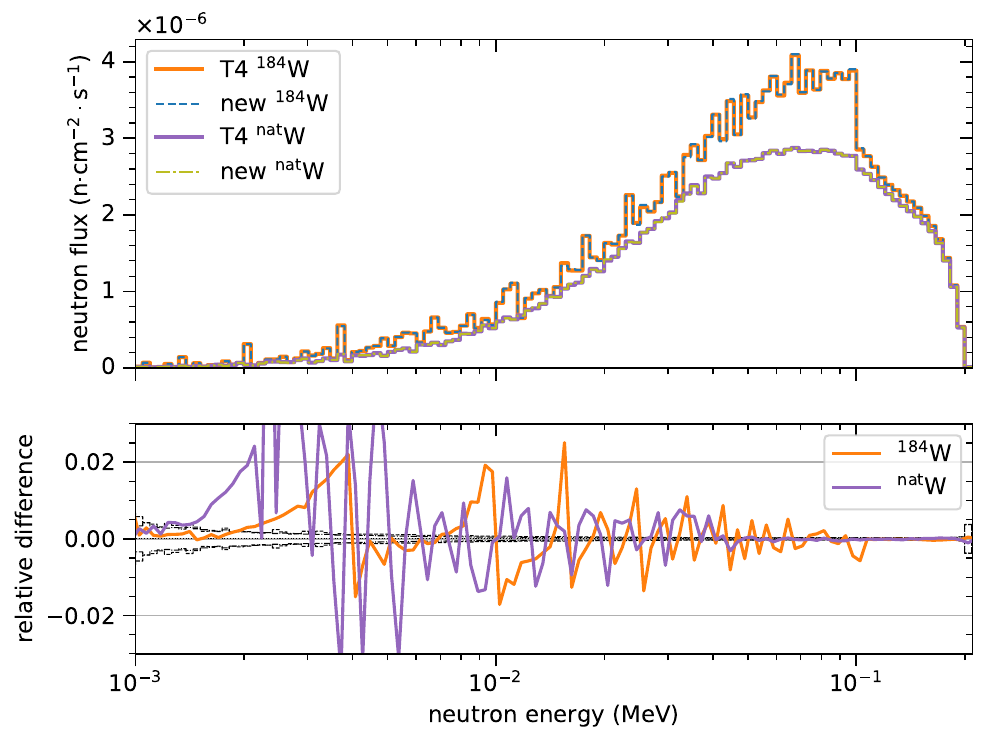}
     \caption{Comparison of the neutron flux obtained with {\geant} for two sphere benchmarks containing $^{184}$W and $^\mathrm{nat}$W  using original {\tripoli} PT libraries from CALENDF and the newly produced ones. This validates the production procedure of PT with CALENDF, namely our new processing code available on the following GitLab \cite{MarekGitLab2024}. In the relative difference plots, the 3$\sigma_{\text{stat}}$ uncertainties are represented by the dashed lines.}     
     \label{fig:T4vsnew}
\end{figure}

Firstly, both codes generate PT for total (MT=1), elastic (MT=2), fission (MT=18) and radiative capture (MT=102) cross sections. While CALENDF also generates PT for the inelastic channels, if any, NJOY uses the smooth inelastic average cross sections.
Secondly, in CALENDF the user can define the energy bin limits and has to make sure that the energy grid is thin enough. Then PT are computed for each energy bin of the grid. Since one PT cross section corresponds to one realization, it will vary for different random generator seeds. This leads to randomly scattered cross section values around the average smooth cross section as can be seen in Figure \ref{fig:w184groups}. The average smooth cross section is recovered when applying a moving average algorithm as shown in Figure \ref{fig:w184groups}. On the other hand, NJOY computes PT in a much smaller number of energy points, at which the averaged cross section reconstructed from the PT and averaged smooth cross section are always equal. During the neutron transport simulation, for a given neutron energy, the corresponding PT cross sections are obtained by linear interpolation between the closest energy points at which PT have been computed. 
To get an accurate URR description a trade-off has to be done between a dense energy bin structure with less samples per bin (4 for ipreci=4 to 16 with ipreci=6) as in CALENDF and less energy points and more samples per point (minimum 16) as in NJOY. The difference between cross sections for different precision criteria in CALENDF can be seen in Figure \ref{fig:w184ipreci}. For $^{184}$W in \geant, Figure \ref{fig:ipreciW184benchmark} shows that the differences can be up to $20$ \% and that the neutron spectrum is smoother with the precision criteria ipreci=6 than with ipreci=4. The cross section calculated with higher precision is less scattered around the smooth one and so is closer to it. 
\newline
The influence of the chosen number of generated ladders in NJOY on the \mcnp simulation results can be seen in Figure \ref{fig:ladders}. Overall increasing the minimum value of 16 ladders to 32 and then to 64 ladders in NJOY shows that the differences are not greater than 5 \% with the studied materials.
\indent
A tool to convert ENDF files to probability table libraries directly usable by \geant has been written and made available to the community on GitLab \cite{MarekGitLab2024}. The parameter settings used in this paper are now detailed for the two processing codes. In NJOY the cross section reconstruction tolerance is set to 0.1 \% to Doppler broaden the cross section with the RECONR and BROADR modules and $\sigma_0$ to 10$^{10}$ barns for an infinite dilution. Moreover the number of probability bins is set to 20 and the number of resonance ladders to 16.
In CALENDF, MTREP have been defined as presented in Table \ref{tab:calendf_njoy_mt}. LCORSCT is set to false, meaning that the total cross-sections is the sum of the partial cross sections and LFORMRF is also set to false, meaning that the formalism to describe RRR is the one recommended by the evaluator, yet this should not impact the PT computed in the URR. 
The energy grid is the one from the example named ``inu238e'' from \cite{Sublet2011} having 11232 energy groups. 
The infinite dilution value is 10$^{10}$ barns as well and the precision parameter is set to ipreci=4 (see above discussion). The PTs for \geant are then prepared from the .tpc file, \textit{i.e.} after regrouping them as recommended in \cite{Sublet2011}, and all negative cross-sections are set to 0. This tool has been successfully validated in comparing its generated PT with the one used by \tripoli as can be seen in Figure~\ref{fig:T4vsnew}. The differences are larger than the statistical uncertainties but they are still less than \hbox{3 \%}, which can be explained by the different CALENDF settings to prepare the data. This tool has not been benchmarked against previous PT from NJOY since no PT data set is distributed with \mcnp. Therefore this tool has been used in this work to prepare PT from NJOY, in ACE format, for both \mcnp and \geant.

\newpage
\clearpage

\subsection{Technical implementation in Neutron-HP package}
\label{sec:implbench}

\begin{figure}[!h]
    \centering
    \includegraphics[width=\textwidth]{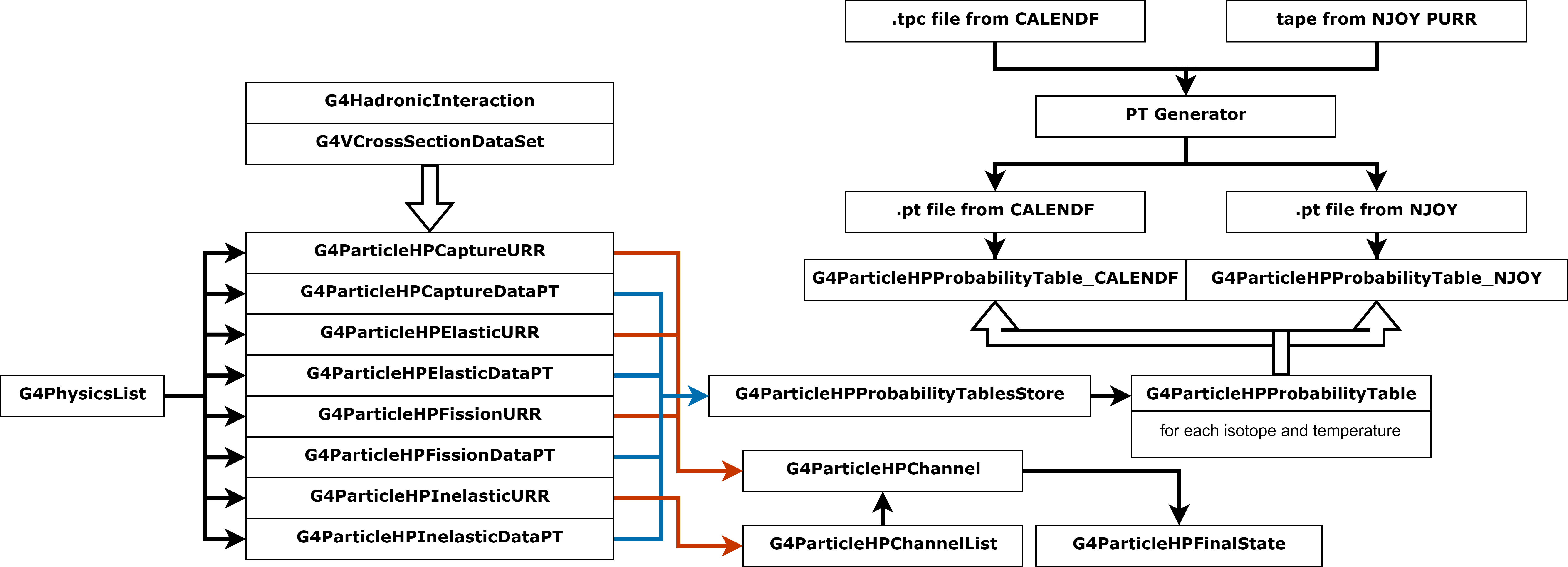}
    \caption{Dependencies of the new classes allowing to handle the treatment of URR with probability tables in the \geant Neutron-HP package.}
    \label{fig:PTimplementation}
\end{figure}

For each reaction cross section \_\emph{XXX}\_ $\in$ \{Elastic, Fission, Capture, Inelastic\} two new classes named G4ParticleHP\_\emph{XXX}\_DataPT and G4ParticleHP\_\emph{XXX}\_URR have been written. These classes leverage the new class G4ParticleHPIsoProbabilityTable, which is created for each isotope and temperature. 
Depending on the PT data either from NJOY or CALENDF, the derived class from G4ParticleHPIsoProbabilityTable, respectively G4ParticleHPIsoProbabilityTable\_NJOY or G4ParticleHPIsoProbabilityTable\_CALENDF, are instantiated. In these classes the probability table file is read and the proper selection of the cross-section is performed. The most important feature to correctly handle probability table data during the simulation is to keep the same random number to sample the cross section in different spatial regions for a given isotope and for a given neutron trajectory having a given energy. An overview of the implementation of the new PT classes in the \geant Neutron-HP package is presented in Figure \ref{fig:PTimplementation}.

\section{Validations of probability tables in \geant}
\label{sec:validations}

The validation procedure is made in three steps. Firstly, results for a \emph{single isotope} sphere benchmark from \mcnp, \tripoli and \geant are compared allowing to validate the good treatment of probability tables in \geant. Secondly, a benchmark with a material having \emph{multiple isotopes} are performed allowing to validate that the URR limits of all isotopes are correctly handled with respect to each other. Thirdly, a case where the sphere is made of one single isotope with \emph{different temperatures} is studied to validate the correct handling of multiple temperatures. Theses different validation steps are now presented.

\newpage
\clearpage

\subsection{Single isotope material}

\begin{figure}[!h]
     \centering
     \begin{subfigure}[b]{0.49\textwidth}
         \centering
         \includegraphics[width=\textwidth]{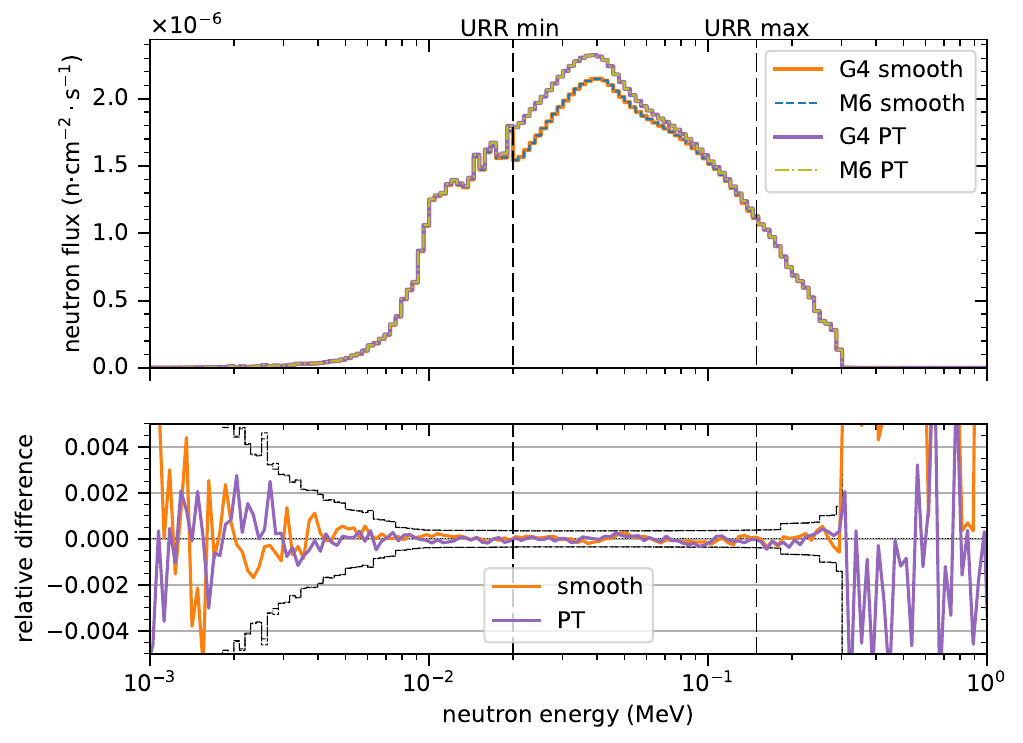}
         \caption{$^{238}$U - \geant \textit{vs.} \mcnp ~-- flux inside the sphere.}
     \end{subfigure}
     \hfill
     \begin{subfigure}[b]{0.49\textwidth}
         \centering
         \includegraphics[width=\textwidth]{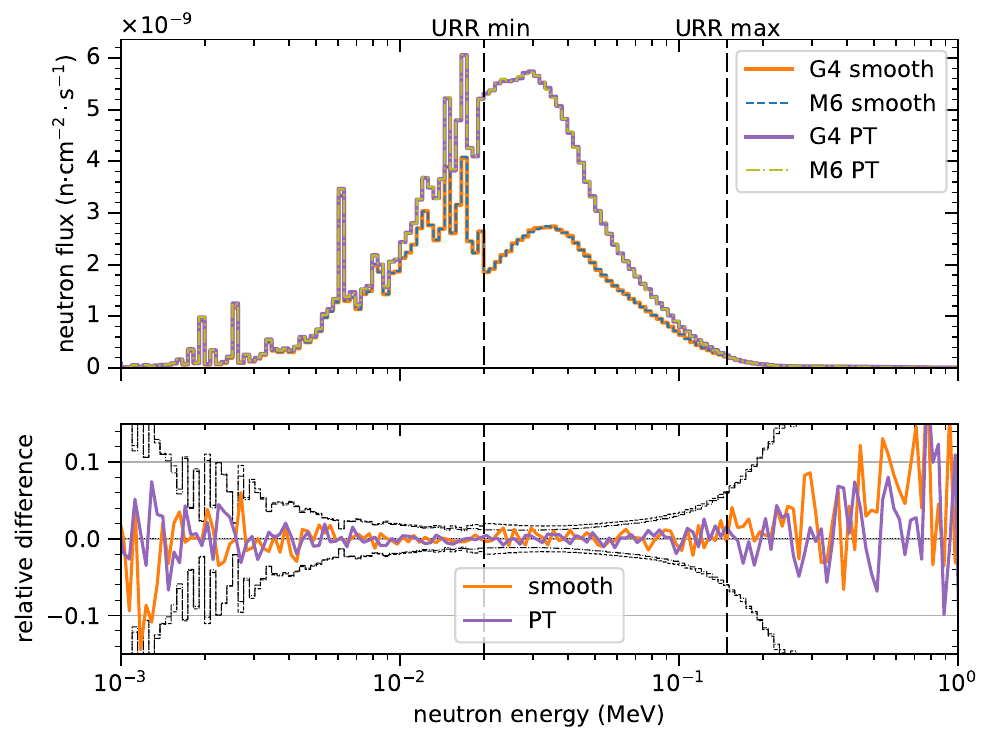}
         \caption{$^{238}$U - \geant \textit{vs.} \mcnp  ~-- outgoing flux.}
     \end{subfigure}
     \begin{subfigure}[b]{0.49\textwidth}
         \centering
         \includegraphics[width=\textwidth]{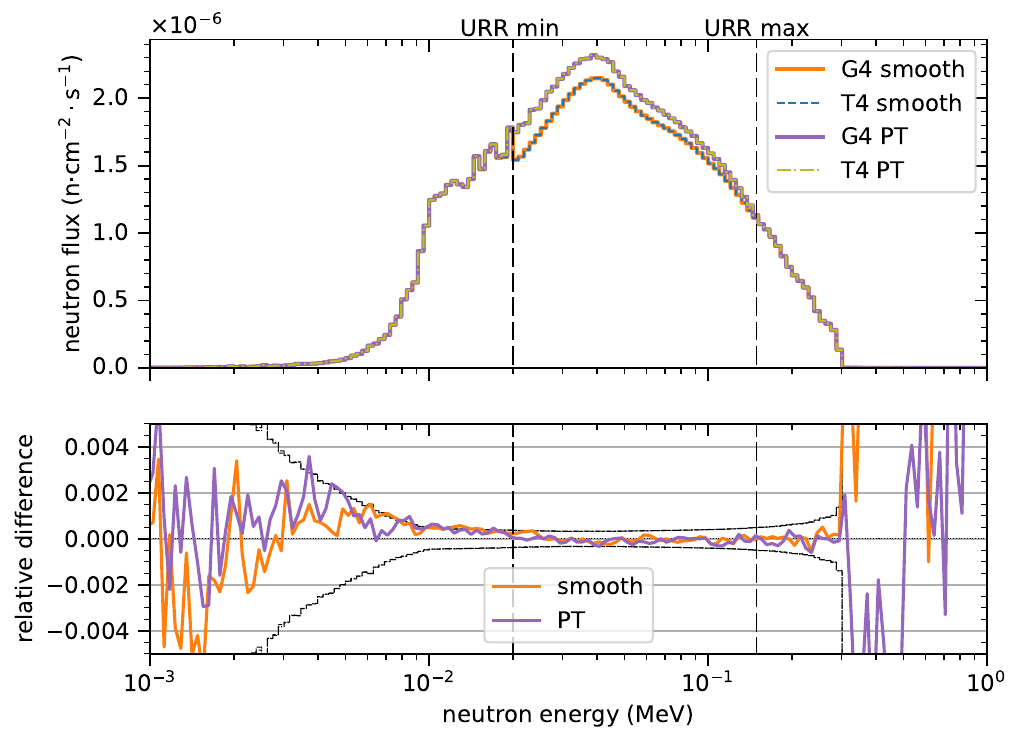}
         \caption{$^{238}$U - \geant \textit{vs.} \tripoli  ~-- flux inside the sphere.}
     \end{subfigure}
      \hfill
     \begin{subfigure}[b]{0.49\textwidth}
         \centering
         \includegraphics[width=\textwidth]{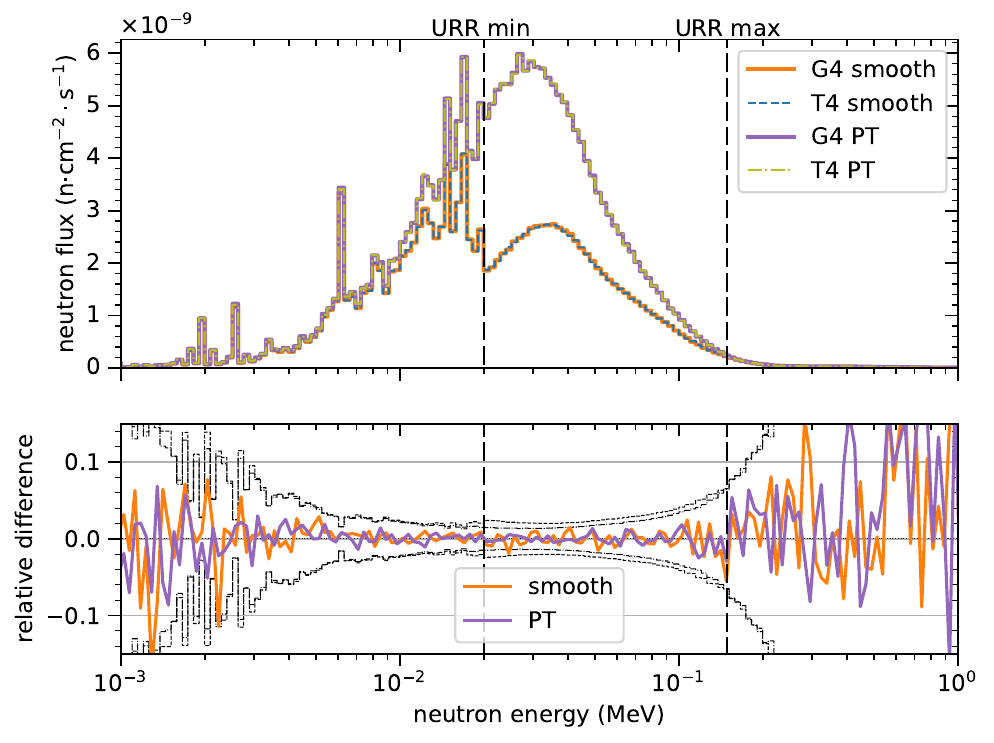}
         \caption{$^{238}$U - \geant \textit{vs.} \tripoli  ~-- outgoing flux.}
     \end{subfigure}
        \caption{Comparison of the neutron fluxes inside (left column) and escaping (right column) the sphere made of $^{238}$U between \geant and \mcnp (upper plots) and between \geant and \tripoli (lower plots).
        Calculations are performed with the use of probability tables (PT) and without (smooth cross section). For each, relative comparisons are made with \mcnp and \tripoli, considered to be the reference. 
        In the relative difference plots, the 3$\sigma_{\text{stat}}$ uncertainties are represented by the dashed lines.}
        \label{fig:uraniumm6t4}
\end{figure}

\begin{figure}[!h]
     \centering
     \begin{subfigure}[b]{0.49\textwidth}
         \centering
         \includegraphics[width=\textwidth]{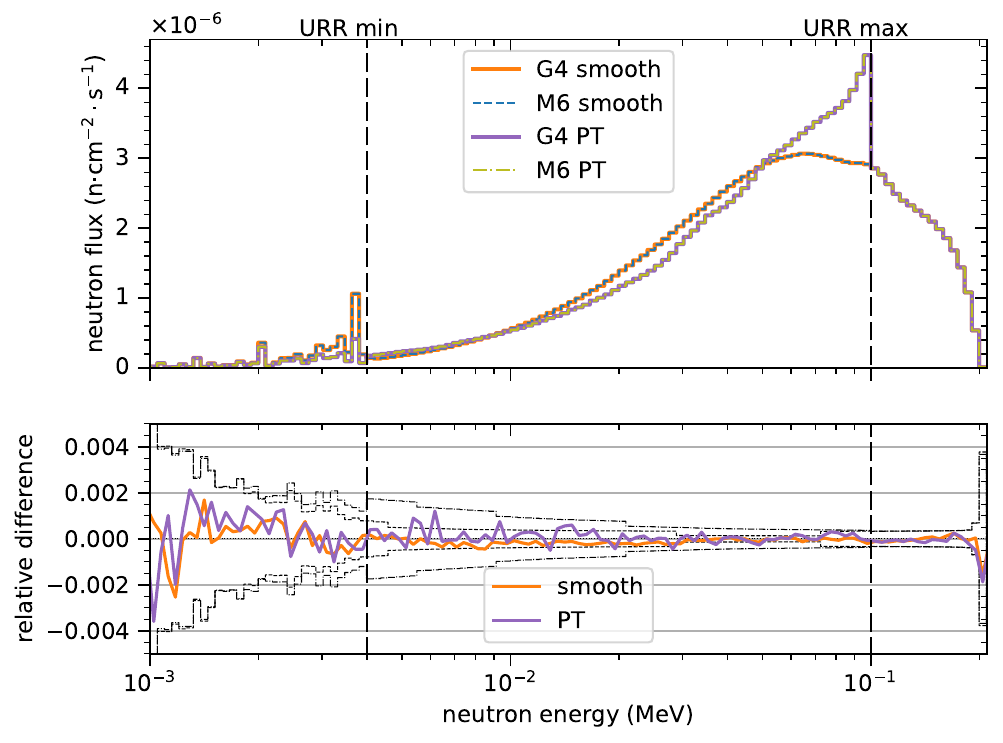}
         \caption{$^{184}$W -  \geant \textit{vs.}  \mcnp  ~-- flux inside the sphere.}
     \end{subfigure}
     \hfill
     \begin{subfigure}[b]{0.49\textwidth}
         \centering
         \includegraphics[width=\textwidth]{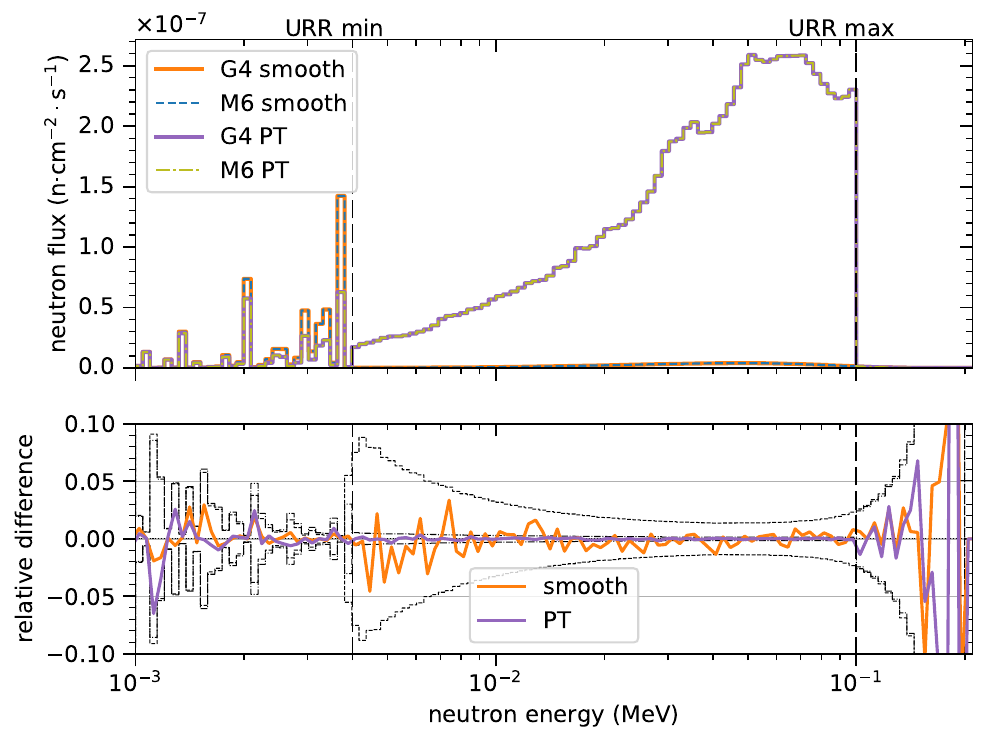}
         \caption{$^{184}$W -  \geant \textit{vs.} \mcnp  ~-- outgoing flux.}
     \end{subfigure}
     \begin{subfigure}[b]{0.49\textwidth}
         \centering
         \includegraphics[width=\textwidth]{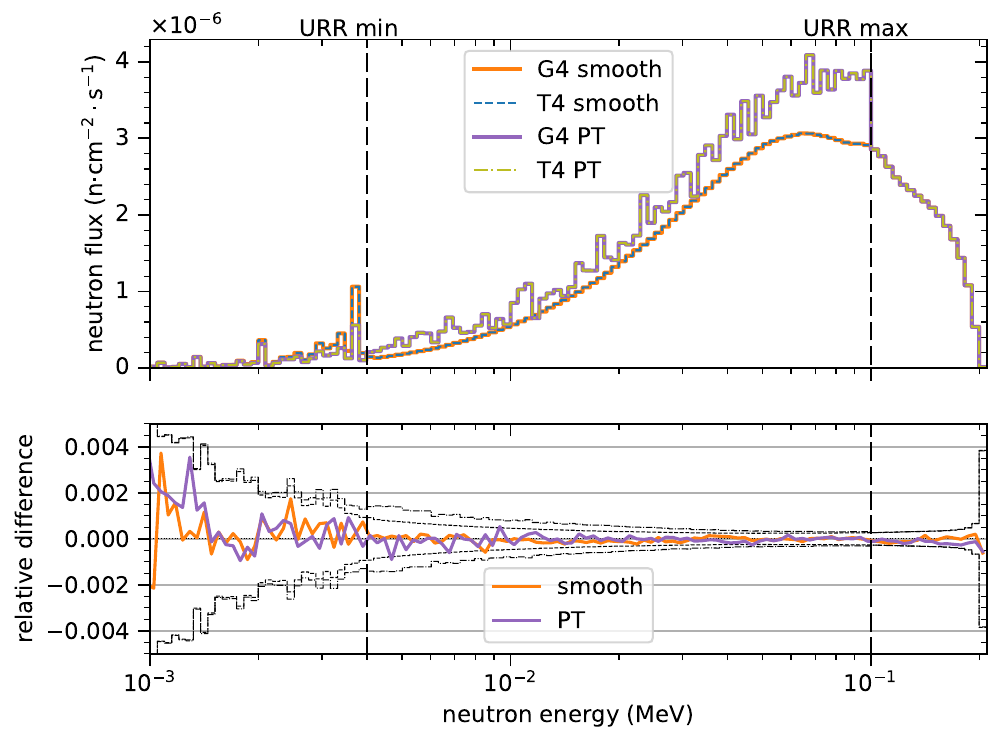}
         \caption{$^{184}$W -  \geant \textit{vs.} \tripoli  ~-- flux inside the sphere.}
     \end{subfigure}
      \hfill
     \begin{subfigure}[b]{0.49\textwidth}
         \centering
         \includegraphics[width=\textwidth]{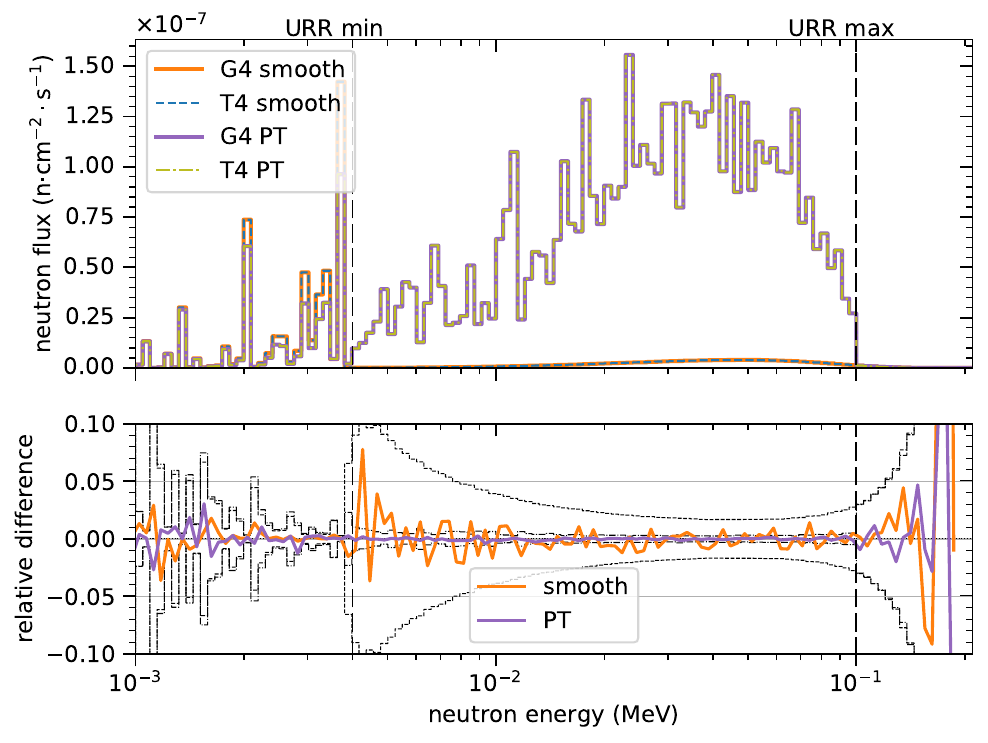}
         \caption{$^{184}$W -  \geant \textit{vs.} \tripoli  ~-- outgoing flux.}
     \end{subfigure}
        \caption{Comparison of the neutron fluxes inside (left column) and escaping (right column) the sphere made of $^{184}$W between \geant and \mcnp (upper plots) and between \geant and \tripoli (lower plots).
        Calculations are performed with the use of probability tables (PT) and without (smooth cross section). For each, relative comparisons are made with \mcnp and \tripoli, considered to be the reference.
        In the relative difference plots, the 3$\sigma_{\text{stat}}$ uncertainties are represented by the dashed lines.}
        \label{fig:tungsten184M6T4}
\end{figure}

This first validation step consists in studying a sphere made of one single isotope allowing to validate the correct treatment of the PT. For that purpose a benchmark with $^{238}$U and $^{184}$W, without and with PT, are performed whose results are respectively presented in Figures \ref{fig:uraniumm6t4} and \ref{fig:tungsten184M6T4}. 
 First of all the comparison of neutron fluxes between \geant and \mcnp or \tripoli shows differences less than the statistical uncertainties. Therefore the probability table method implementation in \geant is validated. It has to be noticed that the $^{184}$W URR upper energy limit at $100$ keV is clearly visible in most cases and gives rather non-physical results caused by the cross section definition in the ENDF file. 
Secondly, for the inside and escaping sphere fluxes, there are large differences between the use of smooth cross section and PT in the URR. 
For example, behind a uranium shielding the outgoing flux increases up to a factor 2 when using PT description of the URR compared to the use of smooth cross section as can be seen in Figure \ref{fig:uraniumm6t4}. These results are in agreement with the one from ref. \cite{Carter1998}. The differences are even larger for the $^{184}$W isotope case (Figure \ref{fig:tungsten184M6T4}) since the outgoing neutron flux is around $15\times$ higher with NJOY PT and $9\times$ higher with CALENDF PT compared to the use of smooth cross sections.

\subsection{Multiple isotope material}

\begin{figure}[!h]
     \centering
     \begin{subfigure}[b]{0.49\textwidth}
         \centering
         \includegraphics[width=\textwidth]{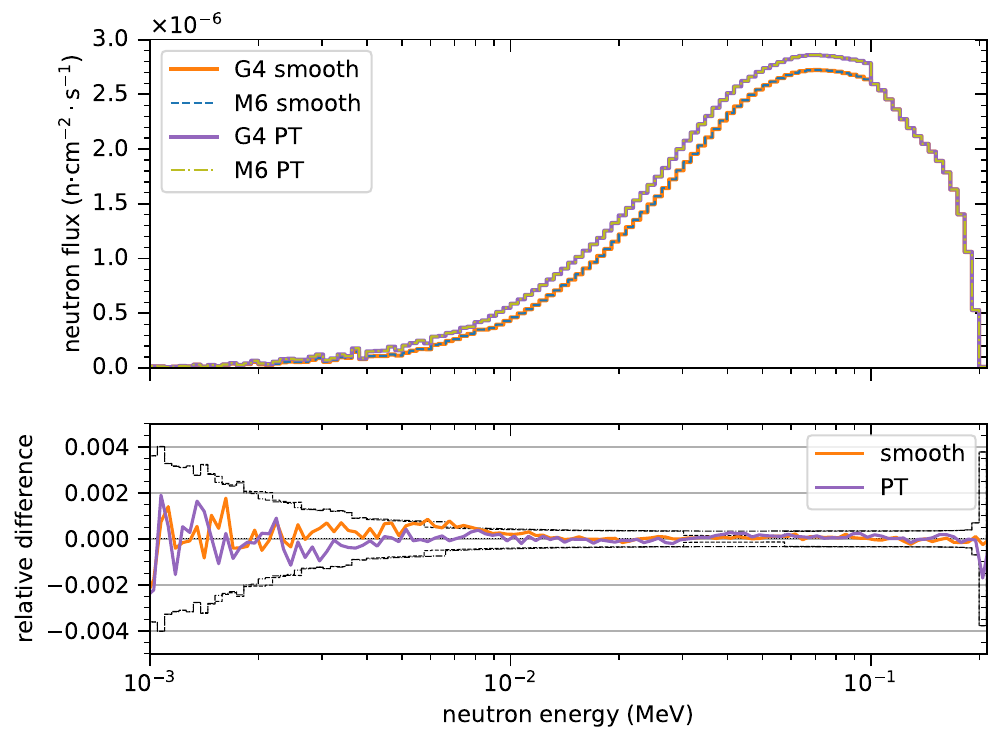}
         \caption{$^\mathrm{nat}$W - \geant \textit{vs.} \mcnp ~--  flux inside the sphere.}
     \end{subfigure}
     \hfill
     \begin{subfigure}[b]{0.49\textwidth}
         \centering
         \includegraphics[width=\textwidth]{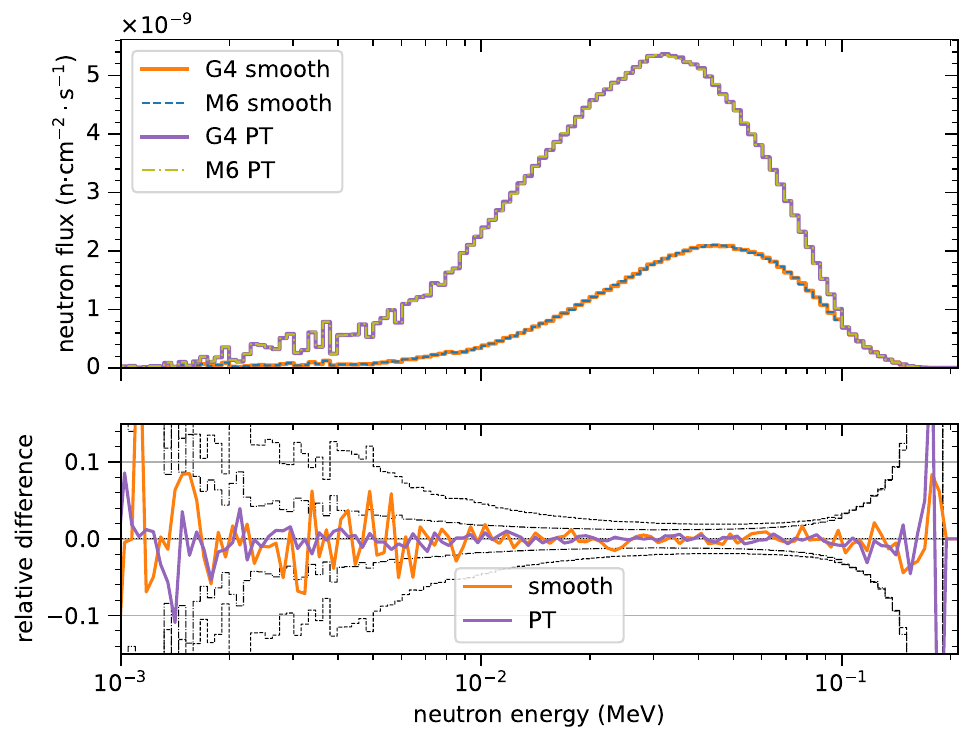}
         \caption{$^\mathrm{nat}$W - \geant \textit{vs.} \mcnp ~--  outgoing flux.}
     \end{subfigure}
     \begin{subfigure}[b]{0.49\textwidth}
         \centering
         \includegraphics[width=\textwidth]{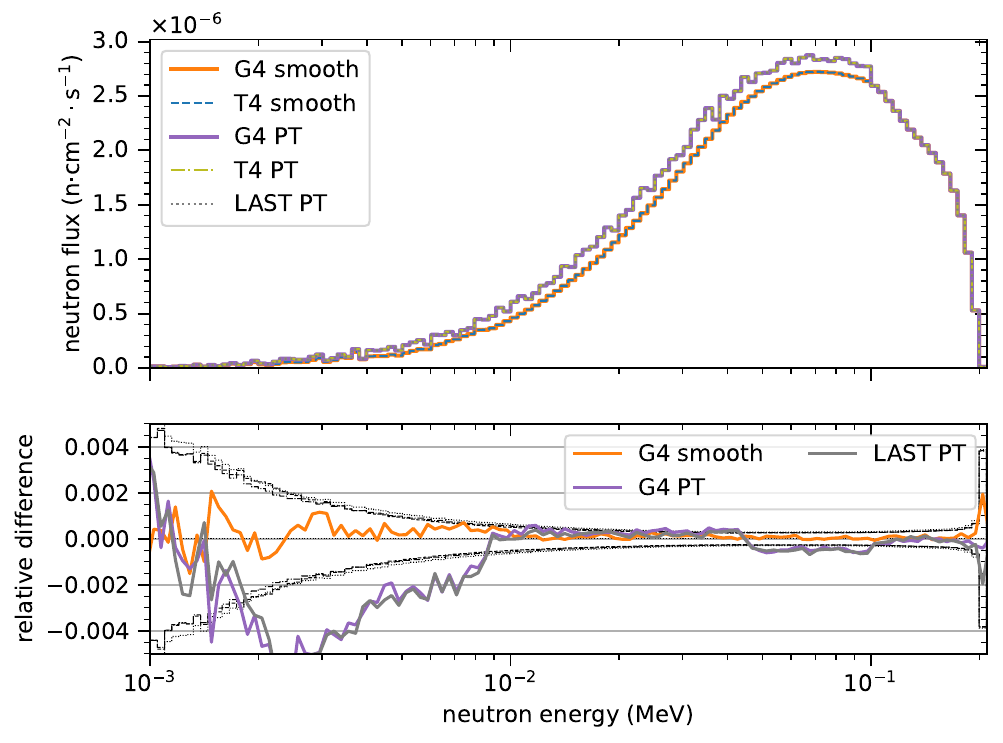}
         \caption{$^\mathrm{nat}$W - \geant, LAST \textit{vs.} \tripoli ~--  flux inside the sphere.}
         \label{fig:G4T4WsphereLAST}
     \end{subfigure}
      \hfill
     \begin{subfigure}[b]{0.49\textwidth}
         \centering
         \includegraphics[width=\textwidth]{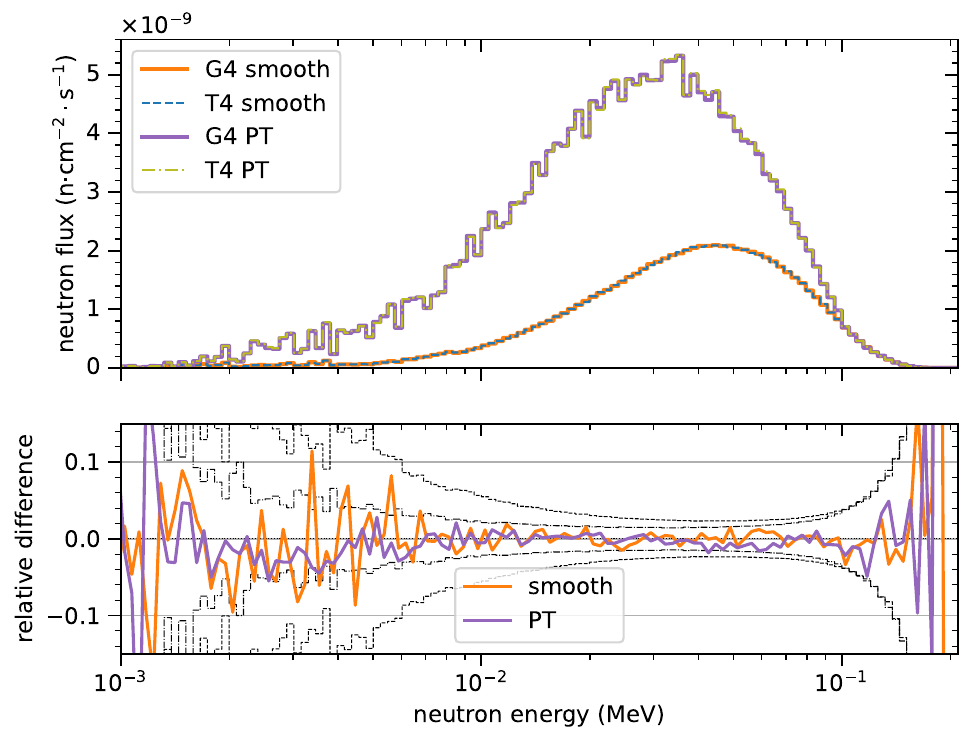}
         \caption{$^\mathrm{nat}$W - \geant \textit{vs.} \tripoli ~--  outgoing flux.}
     \end{subfigure}
        \caption{Comparison of the neutron fluxes inside (left column) and escaping (right column) the sphere made of $^\mathrm{nat}$W between \geant and \mcnp (upper plots) and between \geant and \tripoli (lower plots).
        Calculations are performed with the use of probability tables (PT) and without (smooth cross section). For each, relative comparisons are made with \mcnp and \tripoli, considered to be the reference.
        In the relative difference plots, the 3$\sigma_{\text{stat}}$ uncertainties are represented by the dashed lines.}
        \label{fig:tungstenM6T4}
\end{figure}

This second validation step consists of studying a sphere made of a natural tungsten element, \textit{i.e.} with the different tungsten isotopes with their natural abundances as presented in Table \ref{tab:material}. This allows to validate that isotopes with different URR energy limits are correctly taken into account with respect to each other in the simulation. The results presented in Figure \ref{fig:tungstenM6T4} again show that the difference between simulations without and with PT are significant. \geant and \mcnp agree well with each other within the statistical uncertainties. This validates the correct handling of the different isotopes PT. However when comparing \geant and \tripoli, differences larger than the statistical uncertainties arise. This is clearly visible in the energy ranges $2.2-8.5$~keV (corresponding to the lower limit of the single isotope URR range) and $45-100$~keV (corresponding to the upper limit of the single isotope URR range) mainly for the flux inside the sphere, \textit{i.e.} in the energy range where the cross section from point-wise and PT cross sections are combined together for the different isotopes.

Since the \mcnp and \geant multiple-isotope calculations agree with each other and that all the single tungsten isotope calculations agree with each other for both \mcnp/\geant and \tripoli/\geant as can be seen in the appendix respectively in Figures \ref{fig:tungstenisotopesG4M6} and \ref{fig:tungstenisotopesG4T4}, this indicates that the source of discrepancies could come from the isotope mixing in \tripoli. This trend is also observed with another MC code named LAST \cite{Tamagno2019} that uses the same libraries and input files as \tripoli as can be seen in Figure \ref{fig:G4T4WsphereLAST}. This is still under investigation.

\subsection{Same isotope with different temperatures}

\begin{figure}[!h]
     \centering
     \begin{subfigure}[b]{0.49\textwidth}
         \centering
         \includegraphics[width=\textwidth]{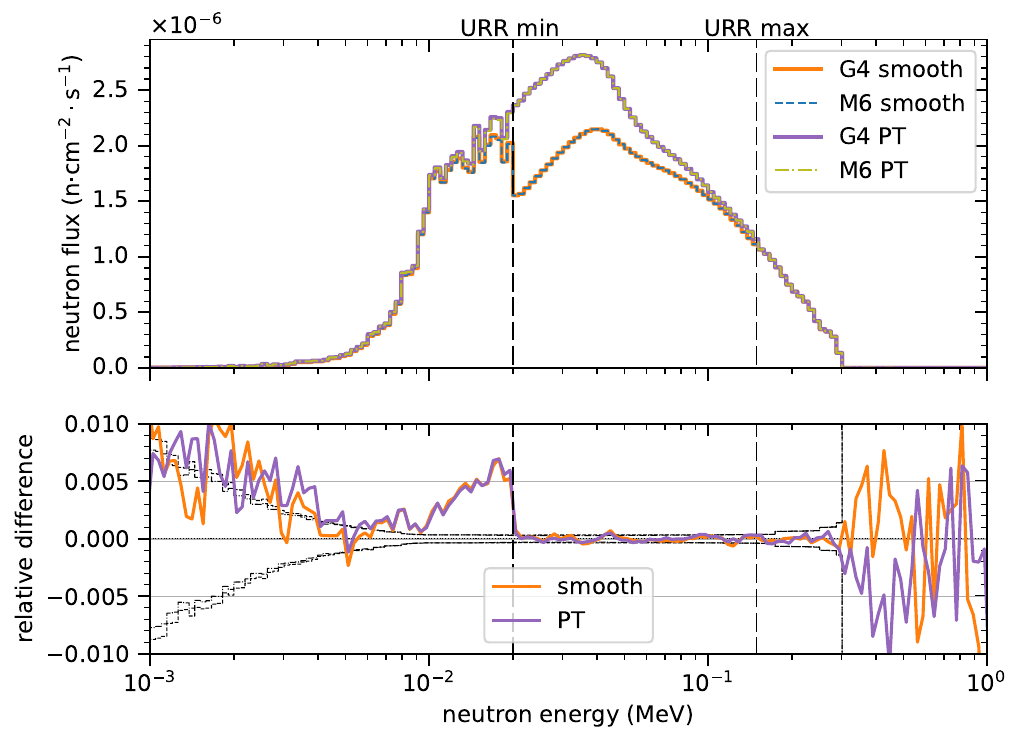}
     \end{subfigure}
     \hfill
     \caption{Comparison of the neutron fluxes inside a hollow sphere made of $^{238}$U at 100 K containing an inner sphere of 25 cm radius at 10 K, between \mcnp and \geant codes with the use of probability tables (PT) and without (smooth cross section). For the relative comparison \mcnp is considered to be the reference. In the relative difference plots, the 3$\sigma_{\text{stat}}$ uncertainties are represented by the dashed lines.}
     \label{fig:Utemp}
\end{figure}

This third validation step consists of a hollow sphere made of a single isotope with 50 cm radius at 100~K containing an inner sphere of 25 cm radius at 10 K. Mixing the two temperatures allows to validate the correct handling of multiple temperatures for the same isotope. These two temperatures have been chosen to be low because in this case only the on-the-fly Doppler broadening of the 0 K cross-section in \geant can be used with a limited computational cost. Indeed the workaround made in \geant to use pre-Doppler broadened cross sections cannot be used here (\textit{cf.} Section \ref{sec:methodology}) since two temperatures are now involved. 
The neutron flux inside the whole sphere is presented in Figure~\ref{fig:Utemp} and exhibits a difference larger than the statistical uncertainty below 20 keV for both cases with and without PT. This hints that it is due to the stochastic on-the-fly Doppler broadening in \geant. In fact, as can be seen in Table \ref{tab:xsDB}, there are discrepancies up to 2.5 \% between the value of the cross section obtained with the SIGMA1 algorithm of NJOY and the average value from the stochastic on-the-fly Doppler broadening algorithm of \geant. Despite these effects, \mcnp and \geant results agree with each other in the URR energy range validating the correct handling of different temperatures of PT for the same isotope. No comparison has been made here with \tripoli because in the official release there are no cross sections at 10 K and 100 K.

\begin{table}[h!]
\centering
\caption{Doppler broadened cross-sections calculated with the SIGMA-1 algorithm from NJOY (exact Doppler broadening) and with the  stochastic on-the-fly (OTF) Doppler broadening algorithm of \geant for $^{238}$U at $19$ keV for 10 K and 100 K temperatures. The cross section from \geant are computed with 1000 samples. The relative difference (Rel. diff.) between NJOY (reference) and \geant is also presented.}

\begin{tabular}{l|lll|lll}
  Channel   & \multicolumn{3}{c|}{Cross section at 10K (barns)}  & \multicolumn{3}{c}{Cross section at 100 K (barns)}         \\
        & \multicolumn{1}{c}{NJOY} & \multicolumn{1}{c}{G4 OTF} & \multicolumn{1}{c|}{Rel. diff.} & \multicolumn{1}{c}{NJOY} & \multicolumn{1}{c}{G4 OTF} & \multicolumn{1}{c}{Rel. diff.} \\ \hline
elastic (x$10^1$) & 1.055 & 1.082 $\pm$0.001  & 2.53\% & 1.053 & 1.080 $\pm$0.003 & 2.50\% \\
capture (x$10^{-3}$) & 4.481 & 4.564 $\pm$0.059 & 1.87\% & 4.736 & 4.809 $\pm$0.194 & 1.55\% \\
fission (x$10^{-8}$)  & 2.388 & 2.401 $\pm$0.0005 & 0.53\% & 2.388 & 2.401 $\pm$0.0005 & 0.53\%
\end{tabular}
\label{tab:xsDB}
\end{table}

\newpage
\section{Comparisons between impact of NJOY and CALENDF probability tables}
\label{sec:comparisons}

\begin{figure}[!h]
     \centering
     \begin{subfigure}[b]{0.49\textwidth}
         \centering
         \includegraphics[width=\textwidth]{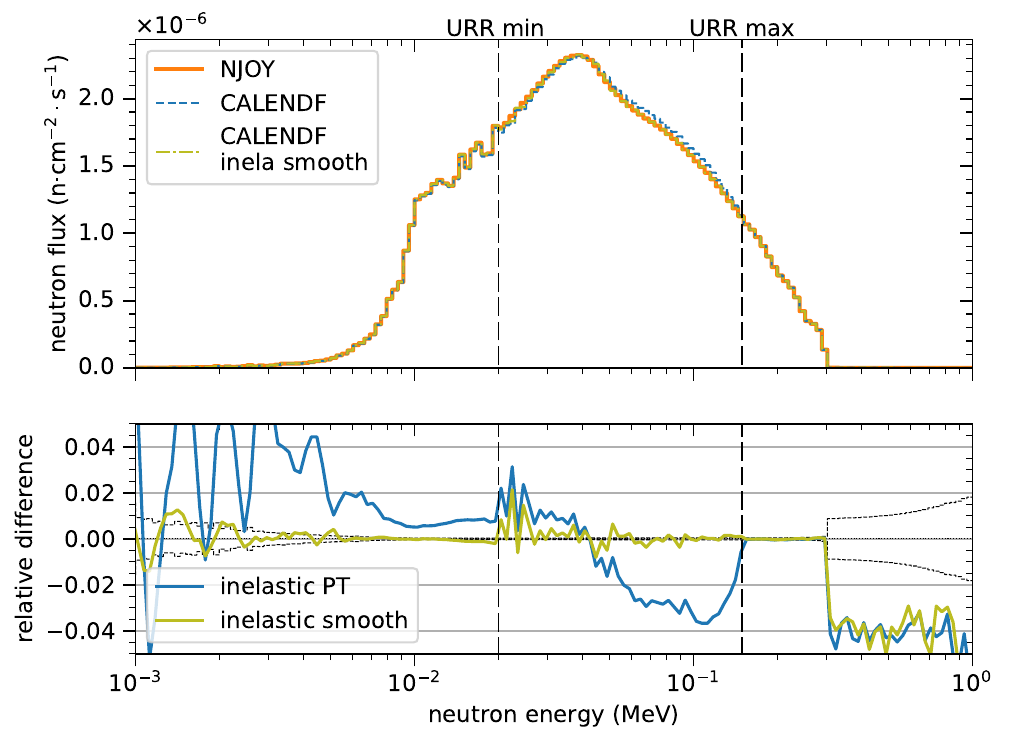}
         \caption{$^{238}$U - \geant - flux inside the sphere}
         \label{fig:njoyvscalendfU}
     \end{subfigure}
     \hfill
     \begin{subfigure}[b]{0.49\textwidth}
         \centering
         \includegraphics[width=\textwidth]{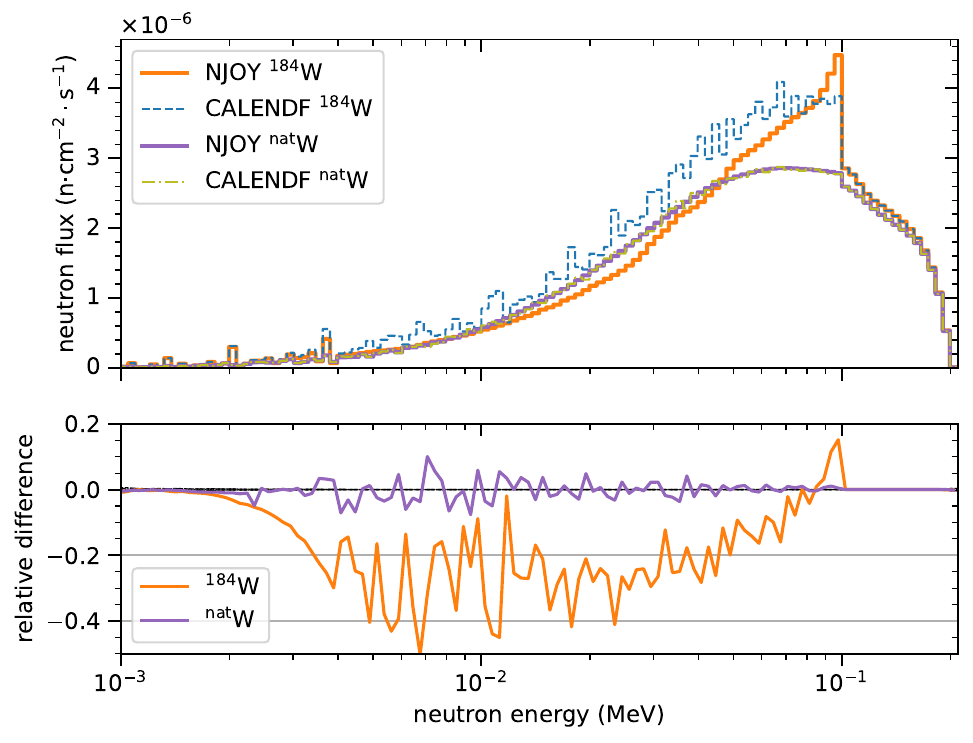}
         \caption{$^{184}$W and $^\mathrm{nat}$W - \geant - flux inside the sphere}
         \label{fig:njoyvscalendfW}
     \end{subfigure}
     \caption{Comparison of the neutron flux inside the sphere using either NJOY and CALENDF probability tables with \geant for $^{238}$U, $^{184}$W and $^{\text{nat}}$W materials. In addition for $^{238}$U, a comparison between the use or not of the inelastic PT from CALENDF is shown. In the relative difference plots, the 3$\sigma_{\text{stat}}$ uncertainties are represented by the dashed lines.}
     \label{fig:njoyvscalendf}
\end{figure}

\geant can now use PT from both NJOY and CALENDF. Therefore it is now a powerful tool to study the impact of the different strategies used during the production of probability tables from URR parameters in each pre-processing code. To our knowledge, this is the first time that this is possible for a neutron transport code \cite{Sublet2009}.  
The comparison between fluxes using NJOY and CALENDF PT can be seen for the neutron flux inside the sphere of $^{238}$U in Figure \ref{fig:njoyvscalendfU} and in Figure \ref{fig:njoyvscalendfW} for $^{184}$W and tungsten element $^{\text{nat}}$W. 
For the three compositions, the choice of the pre-processing code used to generate the PT is shown to have a significant impact, up to $40\%$ for the $^{184}$W case, well above the statistical uncertainty. The differences for $^{\text{nat}}$W and $^{238}$U are below $6$ \% for most energy bins. Additionally the more stochastic nature of the cross section in CALENDF's PT can be seen on the $^{184}$W flux. A deeper investigation of the differences has been performed for the $^{238}$U case.
Calculations have been performed using (blue curve) or not using (green curve) CALENDF's PT for the inelastic reaction. Using smooth inelastic cross section results are in a closer agreement -- withing 2 \% -- with the simulation using NJOY's PT.
The remaining difference on the flux above 300~keV is due to fission neutrons (the initial neutron source spans from 10~keV to 300~keV (see Tab.~\ref{tab:material}).
This may indicate an impact of the PT related to fission which is to first order the main contributor to the fission rate.
Adding PT for the inelastic reaction leads to an overestimation above 45 keV and an underestimation below this energy which corresponds to the (n,n') energy threshold.
This tilt is expected, since in scattering reactions a neutron is moved from an energy to another, and so, even if the estimation of the scattering reaction rate changes, the slowing down process tends to preserve the amount of neutrons.
This shows that the self-shielding of inelastic cross sections is relevant.

\newpage
\clearpage

\section{Conclusion}
\label{sec:conclusion}

The treatment of the Unresolved Resonance Region with the probability table method has been successfully implemented into \geant. The validation performed with the help of a macroscopic benchmark with the reference neutron transport codes \tripoli and \mcnp shows very good agreement within the statistical uncertainties in all cases, except a small discrepancy which is still under investigation and could come from \tripoli when mixing multiple isotopes. With this work \geant can now be used to study the influence of the different strategies used in the pre-processing codes NJOY and CALENDF to generate probability tables which makes, to our knowledge, \geant a unique tool.
This new treatment of the Unresolved Resonance Region with the probability table method will be incorporated in the next \geant release. This makes the Neutron-HP package as precise as other reference neutron transport codes such as \mcnp and \tripoli in terms of physics description. The last drawback of \geant regarding the neutron transport compared to others dedicated codes is its computational time which is still prohibitive, as underlined in this work. This mainly comes from its stochastic on-the-fly Doppler broadening method. In this work, a workaround has been found to use pre-Doppler broadened cross sections from the BROADR module of NJOY, allowing to overcome this last difficulty. It has also been shown that there are small differences between the cross section values obtained with the SIGMA1 and the on-the-fly Doppler broadening, which call for more investigations. Soon the use of pre-Doppler broadened cross sections at a given temperature in \geant will be addressed to speed-up the simulations.

\section*{Acknowledgement}

This work was supported by the French government under the France 2030 program (P2I - Graduate School Physics) under reference ANR-11-IDEX-0003. The presented results were obtained using the CICRR infrastructure, which is financially supported by the Ministry of Education, Youth and Sports - project LM2023041.
The authors wish to sincerely thank Alberto Ribon head of the \geant Hadronic Working Group for fruitful discussions regarding \geant. 
\tripoli~is a registered trademark of CEA.

\appendix

\begin{figure}[htbp]
     \centering
     \begin{subfigure}[b]{0.49\textwidth}
         \centering
         \includegraphics[width=\textwidth]{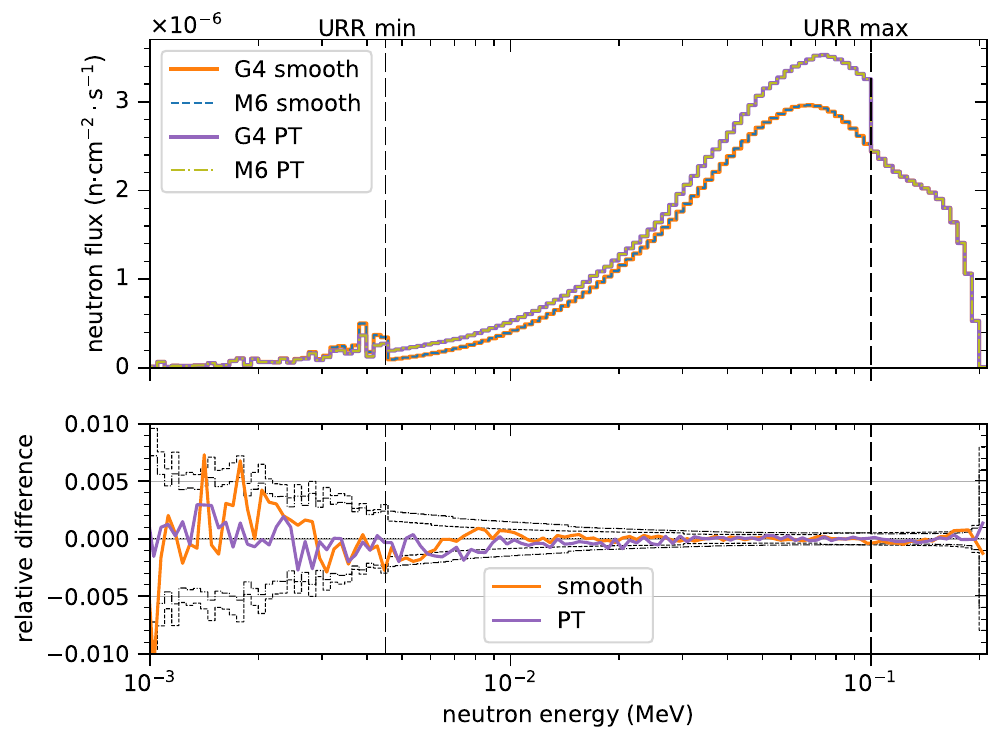}
         \caption{$^{182}$W - flux inside the sphere}
     \end{subfigure}
     \hfill
     \begin{subfigure}[b]{0.49\textwidth}
         \centering
         \includegraphics[width=\textwidth]{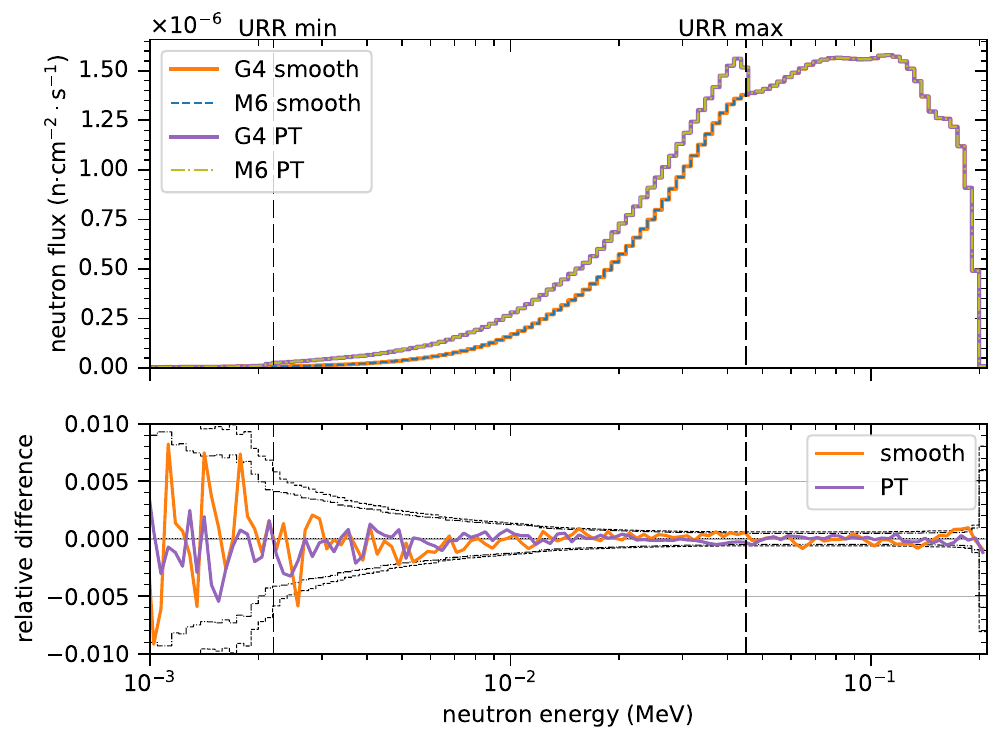}
         \caption{$^{183}$W - flux inside the sphere}
     \end{subfigure}
     \begin{subfigure}[b]{0.49\textwidth}
         \centering
         \includegraphics[width=\textwidth]{G4M6W184sphere.pdf}
         \caption{$^{184}$W - flux inside the sphere}
     \end{subfigure}
      \hfill
     \begin{subfigure}[b]{0.49\textwidth}
         \centering
         \includegraphics[width=\textwidth]{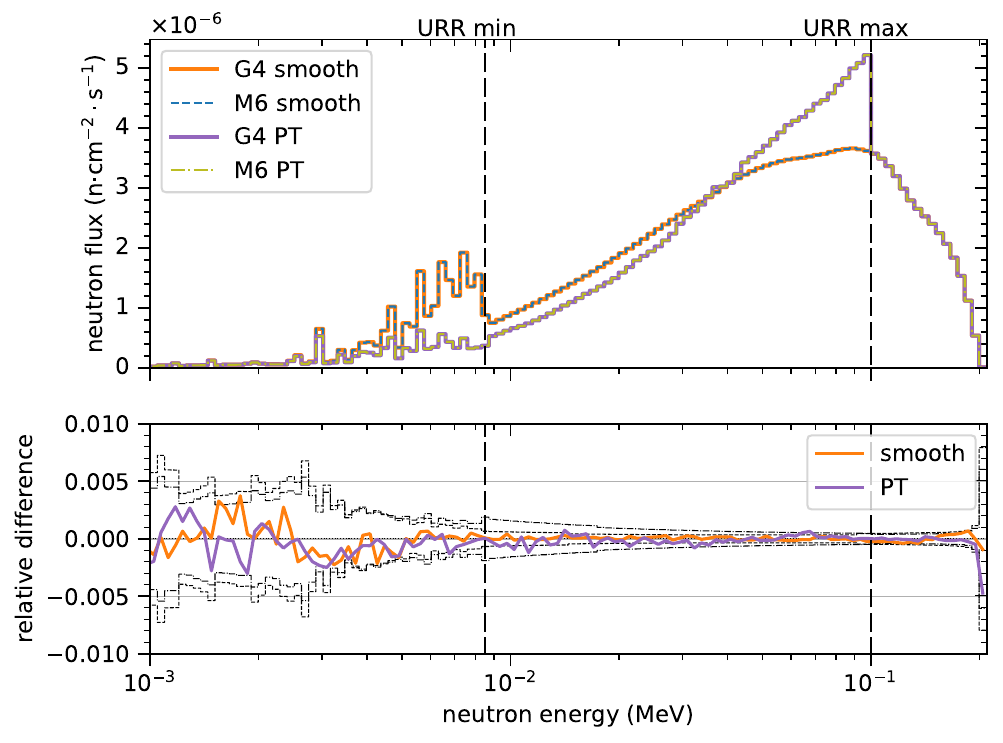}
         \caption{$^{186}$W - flux inside the sphere}
     \end{subfigure}
        \caption{Comparison of the neutron flux inside the sphere between for different tungsten isotopes between \mcnp and \geant codes without (smooth cross section) and with the use of probability tables (PT). For each relative comparison \mcnp are considered to be the reference.}
        \label{fig:tungstenisotopesG4M6}
\end{figure}

\section*{Appendix}
\begin{figure}[htbp]
     \centering
     \begin{subfigure}[b]{0.49\textwidth}
         \centering
         \includegraphics[width=\textwidth]{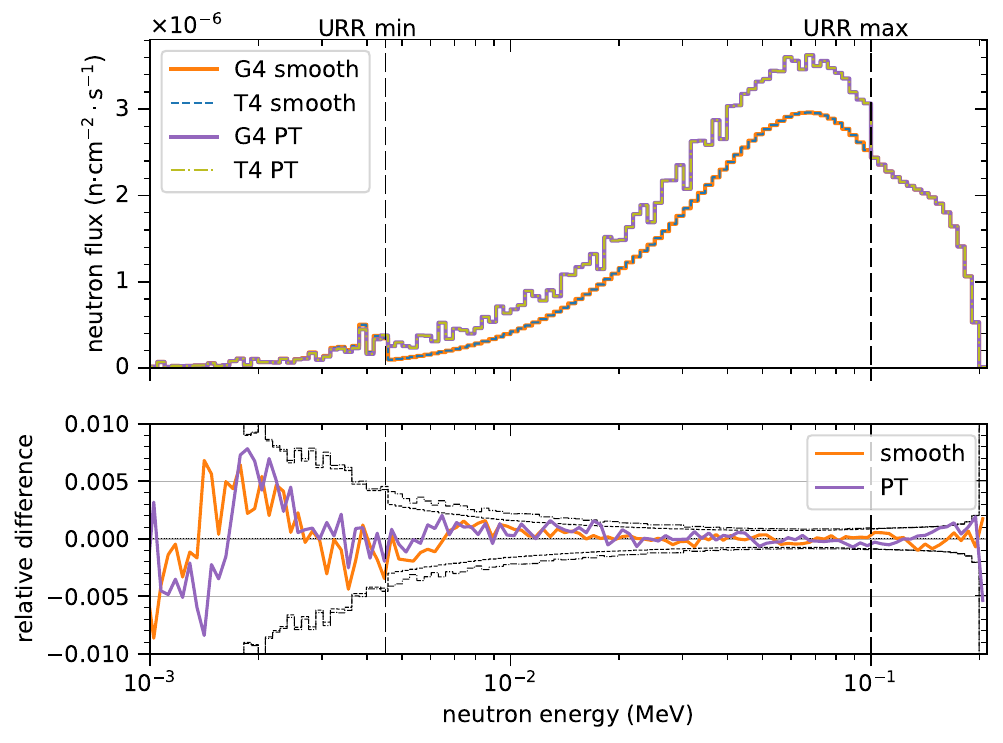}
         \caption{$^{182}$W - flux inside the sphere}
     \end{subfigure}
     \hfill
     \begin{subfigure}[b]{0.49\textwidth}
         \centering
         \includegraphics[width=\textwidth]{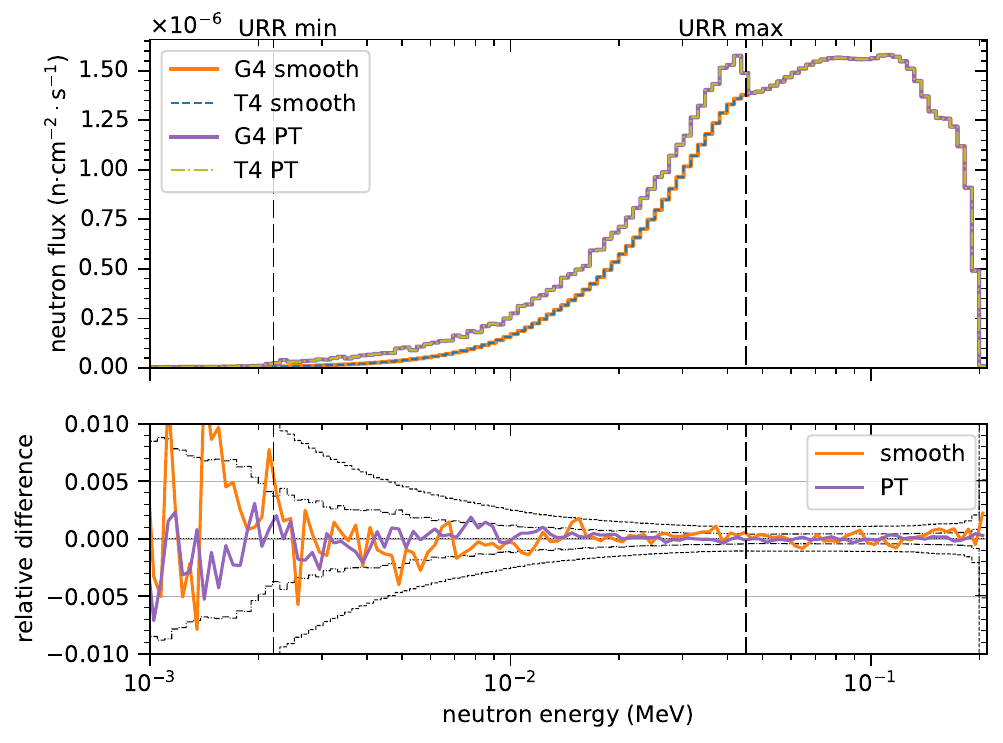}
         \caption{$^{183}$W - flux inside the sphere}
     \end{subfigure}
     \begin{subfigure}[b]{0.49\textwidth}
         \centering
         \includegraphics[width=\textwidth]{G4T4W184sphere.pdf}
         \caption{$^{184}$W - flux inside the sphere}
     \end{subfigure}
      \hfill
     \begin{subfigure}[b]{0.49\textwidth}
         \centering
         \includegraphics[width=\textwidth]{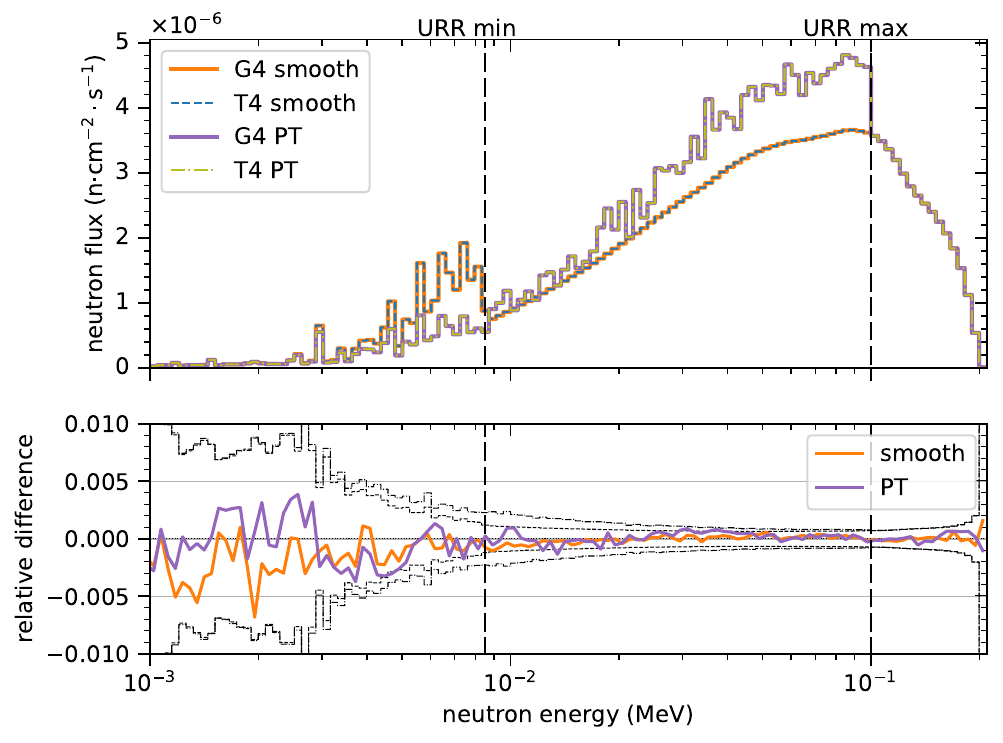}
         \caption{$^{186}$W - flux inside the sphere}
     \end{subfigure}
        \caption{Comparison of the neutron flux inside the sphere between for different tungsten isotopes between \tripoli and \geant codes without (smooth cross section) and with the use of probability tables (PT). For each relative comparison \tripoli are considered to be the reference.}
        \label{fig:tungstenisotopesG4T4}
\end{figure}

\newpage
\bibliographystyle{unsrtnat}
\bibliography{bibliography}

\end{document}